\pdfoutput=1

\documentclass[12pt, draftclsnofoot, onecolumn]{IEEEtran}
\usepackage{amsthm}
\usepackage{multirow}
\usepackage{color}
\usepackage{algorithm}
\usepackage{algpseudocode}
\usepackage{graphicx}
\usepackage{epstopdf}
\newtheorem*{lemma1}{Lemma 1}
\newtheorem*{lemma2}{Lemma 2}
\newtheorem*{lemma3}{Lemma 3}

\newtheorem*{corollary1}{Corollary 1}

\usepackage[T1]{fontenc}
\ifCLASSINFOpdf
\else
\fi
\usepackage{amsmath}
\usepackage{amsmath,bm}
\usepackage{graphicx}
\usepackage{amsmath,amssymb,amsthm,mathrsfs,amsfonts,dsfont}
\usepackage{subfigure}
\begin{document}
%
\title{Auxiliary Beam Pair Enabled AoD and AoA Estimation in Closed-Loop Large-Scale mmWave MIMO Systems}
%
%
%
\author{Dalin Zhu,
        Junil Choi,
        and~Robert~W.~Heath~Jr.
\thanks{Dalin Zhu and Robert W. Heath Jr. are with the Department
of Electrical and Computer Engineering, The University of Texas at Austin, Austin,
TX, 78712 USA, e-mail: \{dalin.zhu, rheath\}@utexas.edu.

Junil Choi is with the Department of Electrical Engineering, Pohang University of Science and Technology (POSTECH), Pohang, Gyeongbuk 37673 Korea, e-mail: junil@postech.ac.kr.

Parts of this work have been presented at IEEE ICASSP 2016 \cite{dz}. This research was partially supported by the U.S. Department of Transportation through the Data-Supported Transportation Operations and Planning (D-STOP) Tier 1 University Transportation Center, and by a gift from Huawei Technologies.}}

\maketitle



\begin{abstract}
Channel estimation is of critical importance in millimeter-wave (mmWave) multiple-input multiple-output (MIMO) systems. Due to the use of large antenna arrays, low-complexity mmWave specific channel estimation algorithms are required. In this paper, an auxiliary beam pair design is proposed to provide high-resolution estimates of the channel's angle-of-departure (AoD) and angle-of-arrival (AoA) for mmWave MIMO systems. By performing an amplitude comparison with respect to each auxiliary beam pair, a set of ratio measures that characterize the channel's AoD and AoA are obtained by the receiver. Either the best ratio measure or the estimated AoD is quantized and fed back to the transmitter via a feedback channel. The proposed technique can be incorporated into control channel design to minimize initial access delay. Though the design principles are derived assuming a high-power regime, evaluation under more realistic assumption shows that by employing the proposed method, good angle estimation performance is achieved under various signal-to-noise ratio levels and channel conditions.
\end{abstract}


%
\IEEEpeerreviewmaketitle

\allowdisplaybreaks

\section{Introduction}

The millimeter-wave (mmWave) band holds promise for providing high data rates in wireless local area network \cite{ieeewlan} and fifth generation (5G) cellular network \cite{jerrypi}-\nocite{rhsp}\cite{fbrh}. The small carrier wavelengths at mmWave frequencies enable synthesis of compact antenna arrays, providing large beamforming gains to enable favorable received signal power \cite{mmwavebook}. Having a large number of antenna arrays, however, makes it difficult to employ fully digital multiple-input multiple-output (MIMO) techniques using one radio frequency (RF) chain per antenna. Instead, hybrid analog and digital precoding has become a means of exploiting both beamforming and spatial multiplexing gains in hardware constrained mmWave cellular systems \cite{molis}-\nocite{wonil}\nocite{omar2}\nocite{liang}\nocite{ahmedwb}\cite{uva}.

Channel knowledge is critical to exploit the full benefit of MIMO techniques in mmWave cellular systems. Classical channel estimation techniques developed for lower-frequency MIMO systems, however, are not applicable for mmWave MIMO due to the use of large antenna arrays, hybrid precoding, and the sparsity of mmWave channels \cite{rhsp}. MmWave specific channel estimation techniques have been proposed in \cite{ahmedce}-\nocite{gcasc}\nocite{hlys}\nocite{rmrconf}\nocite{rmr}\nocite{rmrconf1}\nocite{dghiw}\nocite{covariance}\nocite{singh2}\nocite{wang}\cite{hur}. In \cite{ahmedce}, algorithms that exploit channel sparsity were developed to leverage compressed sensing to perform channel estimation. In \cite{gcasc}, temporal channel correlations were exploited to develop low-complexity compressed sensing algorithms to estimate the channel's coefficients by leveraging the angular domain sparsity of the mmWave channels. In \cite{hlys}, the least square estimation and sparse message passing algorithm were jointly employed in an iterative manner to detect and recover the non-zero entries of the sparse mmWave channels. An open-loop channel estimation strategy was proposed in \cite{rmrconf,rmr}, in which the estimation algorithm was independent of the hardware constraints and applied to either phase shifter or switching networks. The proposed algorithms in \cite{rmrconf,rmr} also exploited the angular sparsity of the mmWave channels and incorporated the hybrid architecture. In \cite{rmrconf1}, compressed measurements obtained from the mmWave channels were exploited to estimate the second order statistics of the channel to enable adaptive multi-user hybrid precoding. In \cite{dghiw}, a support (the index set of non-zero entries in a sparse vector) detection-based channel estimation algorithm was developed for mmWave systems with lens antennas. By determining the supports of all channel components in a successive cancellation fashion, the non-zero channel entries were estimated. The estimation performance of the proposed algorithm in \cite{dghiw}, however, was subject to the estimation error propagation. In \cite{covariance}, a minimum-mean-squared-error (MMSE) hybrid analog and digital channel estimator was developed without exploiting channel sparsity and directional beamforming. In \cite{singh2}, a grid-of-beams (GoB) based approach was proposed to obtain the channel's angle-of-departure (AoD) and angle-of-arrival (AoA). Via exhaustive or sequential search, the best combinations of analog transmit and receive beams, which characterize channel's AoD and AoA, were obtained. A large amount of training is required though, which may be computationally prohibitive and also a source of overhead. Similar ideas of forming beams grids were also investigated in \cite{wang,hur}, though these two papers mainly focused on the hierarchical beam codebook design. Only quantized angle estimation with limited resolution can be achieved via the compressed sensing \cite{ahmedce}-\nocite{gcasc}\nocite{hlys}\nocite{rmrconf}\nocite{rmr}\nocite{rmrconf1}\cite{dghiw} and grid-of-beams \cite{singh2}-\nocite{wang}\cite{hur} based methods. With a small codebook, the resolution of angle estimation becomes low.

Many high resolution subspace based angle estimation algorithms such as MUSIC \cite{music}, ESPRIT \cite{esprit} and their variants \cite{twode} have been of great research interest to the array processing community for decades. Their applications to massive MIMO or full-dimension MIMO to estimate the two-dimensional angles were extensively investigated in \cite{twba}-\nocite{ahtl}\nocite{awll}\nocite{lcycw}\cite{rsll}. For the mmWave frequency band, the MUSIC algorithm was employed for initial user discovery in \cite{vrj} by exploiting mmWave channels' sparsity via directional beamforming. A relatively large number of snap-shots (samples) are required in the subspace based angle estimation algorithms employed in \cite{twba}-\nocite{ahtl}\nocite{awll}\nocite{lcycw}\nocite{rsll}\cite{vrj} to obtain accurate received signal covariance matrix, which in turn, results in high training overhead. Further, it is difficult to directly extend the MUSIC and ESPRIT-type estimators to mmWave systems with the hybrid architecture. This is because with the hybrid architecture, only a reduced-dimension channel matrix can be accessed through the lens of a limited number of RF chains, which makes the estimation of the full MIMO channel difficult.

In this paper, we develop an algorithm for estimating the channel's AoD and AoA through auxiliary beam pair (ABP) design with high accuracy and low training overhead. Pairs of beams were previously employed in monopulse radar systems to improve the estimation accuracy of the direction of arrival \cite{radar0}-\nocite{radar1}\nocite{radaranalysis}\nocite{sgan}\cite{phd}. In amplitude monopulse radar, a sum beam and a difference beam form a pair such that the difference beam steers a null at the boresight angle of the sum beam. By comparing the relative amplitude of the pulse in the pair of two beams, the direction of the object can be determined with accuracy dependent on the received signal-to-noise ratio (SNR).  In our previous work \cite{dz}, the idea of beam pair design was exploited to estimate mmWave communications channels. The proposed approach, however, can only estimate the AoD of a single-path channel assuming omni-directional receive antenna. In this paper, we consider implementing the well structured beam pairs to help acquire high-resolution AoD and AoA estimates under various channel conditions. The main contributions of the paper are:
\begin{itemize}
  \item \emph{Joint AoD and AoA estimation via auxiliary beam pair design}. Both the transmitter and receiver form custom designed analog auxiliary beam pairs to cover given angular ranges. By performing amplitude comparison on the transmit auxiliary beam pairs, a set of ratio measures that characterize the AoD are obtained by the receiver. Similarly, by conducting amplitude comparison on the receive auxiliary beam pairs, the AoA is also estimated at the receiver. Detailed auxiliary beam pair design procedures for both single-path and multi-path channels are illustrated.
  \item \emph{Quantization and feedback options}. The receiver can either quantize the ratio measure that characterizes the AoD or the estimated transmit spatial frequency. We show that quantizing the ratio measure gives better quantization performance than quantizing the estimated transmit spatial frequency, although the difference is marginal.
  \item \emph{Multi-path estimation using multiple RF chains}. Building on the single-path solution, we propose an algorithm for multi-path AoD and AoA estimation employing multiple transmit and receive RF chains. The associated auxiliary beam pair based channel probing matrix is specifically designed for multi-path estimation.
\end{itemize}
The numerical results show that the proposed algorithm is capable of providing high-resolution AoD and AoA estimation under various SNR levels and channel conditions.

The rest of this paper is organized as follows. Section II describes the system model along with a brief discussion on the problem formulation. Section III specifies the design principles of the proposed auxiliary beam pair in estimating both single-path and multi-path channels, and the key differences from the beam pair design in monopulse radar systems. The applications of the proposed design approach to control channel beamforming and multi-user scenario are demonstrated in Section IV, and Section V shows numerical results to validate the effectiveness of the proposed technique. Conclusions are drawn in Section VI.

\textbf{Notations}: $\bm{A}$ is a matrix; $\bm{a}$ is a vector; $a$ is a scalar; $(\cdot)^{\mathrm{T}}$ and $(\cdot)^{*}$ denote transpose and conjugate transpose; $\|\bm{A}\|_{\mathrm{F}}$ is the Frobenius norm of $\bm{A}$ and $\det(\bm{A})$ is its determinant; $\left[\bm{A}\right]_{:,j}$ is the $j$-th column of $\bm{A}$; $\left[\bm{A}\right]_{i,j}$ is the $(i,j)$-th entry of $\bm{A}$; $\mathrm{tr}(\bm{A})$ is the trace of $\bm{A}$. $\textrm{diag}\left(\bm{a},\bm{b}\right)$ is a block diagonal matrix formed with $\bm{a}$ and $\bm{b}$; $\bm{I}_{N}$ is the $N\times N$ identity matrix; $\bm{0}_{N}$ denotes the $N\times 1$ vector whose entries are all zeros; $\mathcal{CN}(\bm{a},\bm{A})$ is a complex Gaussian vector with mean $\bm{a}$ and covariance $\bm{A}$; $\mathbb{E}[\cdot]$ is used to denote expectation.

\section{System Model and Assumptions}
\begin{figure}
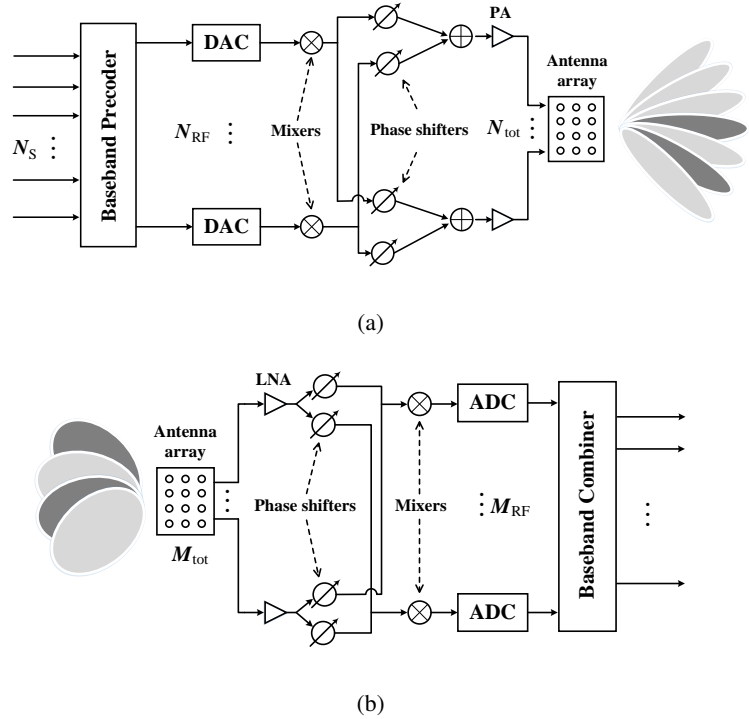

\centering
\subfigure[]{%
\includegraphics[width=4.0in]{transmitter_shared.pdf}
\label{fig:subfigure1}}
\quad
\subfigure[]{%
\includegraphics[width=3.6in]{receiver_shared.pdf}
\label{fig:subfigure2}}
\caption{(a) Shared-array architecture at the transmitter with $N_{\mathrm{RF}}$ RF chains and a total number of $N_{\mathrm{tot}}$ antenna elements. $N_{\mathrm{S}}$ data streams are transmitted. (b) Shared-array architecture at the receiver with $M_{\mathrm{RF}}$ RF chains and a total number of $M_{\mathrm{tot}}$ antenna elements.}
\label{fig:figure}
\end{figure}
We consider a narrowband MIMO system with a hybrid precoded transceiver structure as shown in Fig.~1. A transmitter equipped with $N_{\mathrm{tot}}$ transmit antennas and $N_{\mathrm{RF}}$ RF chains transmits $N_{\mathrm{S}}$ data streams to a receiver equipped with $M_{\mathrm{tot}}$ receive antennas and $M_{\mathrm{RF}}$ RF chains. Here, $N_{\mathrm{S}}\leq M_{\mathrm{RF}}\leq N_{\mathrm{RF}}$, and both the transmitter and receiver are equipped with shared-array antenna architectures \cite{jazhang}. Note that the proposed approach in Section III can be applied to sub-array antenna architecture, with some modifications. As can be seen from Fig.~1, in a shared-array architecture, all antenna elements are jointly controlled by all RF chains sharing the same network of phase shifters. Denote by $\bm{x}=\left[x_{1},\cdots,x_{N_{\mathrm{S}}}\right]^{\mathrm{T}}$ the $N_{\mathrm{S}}\times 1$ vector of symbols such that $\mathbb{E}\left[|x_{k}|^{2}\right]=1$ for $k=1,\cdots,N_{\mathrm{S}}$, and denote by $\bm{y}=\left[y_{1},\cdots,y_{N_{\mathrm{S}}}\right]^{\mathrm{T}}$ the $N_{\mathrm{S}}\times 1$ vector of symbols received across the receive antennas after analog and baseband combining,
\begin{equation}
\bm{y}=\bm{W}_{\mathrm{BB}}^{*}\bm{W}_{\mathrm{RF}}^{*}\bm{H}\bm{F}_{\mathrm{RF}}\bm{F}_{\mathrm{BB}}\bm{x}+\bm{W}_{\mathrm{BB}}^{*}\bm{W}_{\mathrm{RF}}^{*}\bm{n},
\end{equation}
where $\bm{n} \sim \mathcal{CN}(\bm{0}_{M_{\mathrm{tot}}},\sigma^{2}\bm{I}_{M_{\mathrm{tot}}})$ is a noise vector, $\sigma^2=1/\gamma$, and $\gamma$ represents the target SNR. $\bm{F}_{\mathrm{RF}}$ is the $N_{\mathrm{tot}}\times N_{\mathrm{RF}}$ analog precoding matrix at the transmitter. As the analog precoder is implemented using analog phase shifters, $\left[\left[\bm{F}_{\mathrm{RF}}\right]_{:,\jmath}\left[\bm{F}_{\mathrm{RF}}\right]_{:,\jmath}^{*}\right]_{i,i}=\frac{1}{N_{\mathrm{tot}}}$ is satisfied, where $\jmath=1,\cdots,N_{\mathrm{RF}}$ and $i=1,\cdots,N_{\mathrm{tot}}$, i.e., all elements of $\bm{F}_{\mathrm{RF}}$ have equal norm. $\bm{F}_{\mathrm{BB}}$ is the $N_{\mathrm{RF}}\times N_{\mathrm{S}}$ digital baseband precoding matrix at the transmitter such that $\left\|\bm{F}_{\mathrm{RF}}\bm{F}_{\mathrm{BB}}\right\|_{\mathrm{F}}^{2}=1$, $\bm{W}_{\mathrm{BB}}$ and $\bm{W}_{\mathrm{RF}}$ denote $M_{\mathrm{RF}}\times N_{\mathrm{S}}$ and $M_{\mathrm{tot}}\times M_{\mathrm{RF}}$ baseband and analog combining matrices such that $\left\|\bm{W}_{\mathrm{BB}}^{*}\bm{W}_{\mathrm{RF}}^{*}\right\|_{\mathrm{F}}^{2}=1$ and all elements of $\bm{W}_{\mathrm{RF}}$ have equal norm. $\bm{H}$ represents the $M_{\mathrm{tot}}\times N_{\mathrm{tot}}$ narrowband MIMO channel. In this paper, we employ a ray-cluster based spatial channel model which was previously studied in \cite{xu}. Denote by $g_{\ell}$, $\phi_{\ell}$ and $\theta_{\ell}$ the complex path gain, AoA and AoD of path-$\ell$, $N_{\mathrm{p}}$ the total number of paths in the channel, and $\bm{a}_{\mathrm{r}}(\cdot)$ and $\bm{a}_{\mathrm{t}}(\cdot)$ the array response vectors for the receive and transmit antenna arrays. The narrowband channel is therefore represented as
\begin{eqnarray}
\bm{H}=\sqrt{N_{\mathrm{tot}}M_{\mathrm{tot}}}\sum_{\ell=1}^{N_{\mathrm{p}}}g_{\ell}\bm{a}_{\mathrm{r}}(\phi_{\ell})\bm{a}_{\mathrm{t}}^{*}(\theta_{\ell}).\label{chm}
\end{eqnarray}
In this paper, a uniform linear array (ULA) is employed at both the transmitter and receiver to explain the key idea of auxiliary beam pair design. The proposed design approach in Section III can be extended to uniform planar arrays, but this is beyond the scope of our work. Denote by $\lambda$ the wavelength corresponding to the operating carrier frequency, and $d_{\mathrm{t}}$ the inter-element distance of the transmit antenna elements, we have
\begin{equation}
\bm{a}_{\mathrm{t}}(\theta_{\ell})=\frac{1}{\sqrt{N_{\mathrm{tot}}}}\left[1, e^{j\frac{2\pi}{\lambda}d_{\mathrm{t}}\sin(\theta_{\ell})},\cdots, e^{j\frac{2\pi}{\lambda}\left(N_{\mathrm{tot}}-1\right)d_{\mathrm{t}}\sin(\theta_{\ell})} \right]^{\mathrm{T}}.
\end{equation}
Similarly, assuming $d_{\mathrm{r}}$ as the inter-element distance between the receive antenna elements,
\begin{equation}
\bm{a}_{\mathrm{r}}(\phi_{\ell})=\frac{1}{\sqrt{M_{\mathrm{tot}}}}\left[1, e^{j\frac{2\pi}{\lambda}d_{\mathrm{r}}\sin(\phi_{\ell})},\cdots, e^{j\frac{2\pi}{\lambda}\left(M_{\mathrm{tot}}-1\right)d_{\mathrm{r}}\sin(\phi_{\ell})} \right]^{\mathrm{T}}.
\end{equation}
The channel model in (\ref{chm}) can be further expressed as $\bm{H}=\sqrt{N_{\mathrm{tot}}M_{\mathrm{tot}}}\bm{A}_{\mathrm{r}}\textrm{diag}(\bm{g})\bm{A}^{*}_{\mathrm{t}}$ \cite{ahmedce}, where $\bm{g}=\left[g_{1},g_{2},\cdots,g_{N_{\mathrm{p}}}\right]^{\mathrm{T}}$, $\bm{A}_{\mathrm{t}}=\left[\bm{a}_{\mathrm{t}}(\theta_{1}),\bm{a}_{\mathrm{t}}(\theta_{2}),\cdots,\bm{a}_{\mathrm{t}}(\theta_{N_{\mathrm{p}}})\right]$ and $\bm{A}_{\mathrm{r}}=\big[\bm{a}_{\mathrm{r}}(\phi_{1}),\bm{a}_{\mathrm{r}}(\phi_{2}),\cdots,\\\bm{a}_{\mathrm{r}}(\phi_{N_{\mathrm{p}}})\big]$ contain the transmit and receive array response vectors.
\section{Auxiliary Beam Pair Based Channel Estimation}
\begin{figure}
\centering
\includegraphics[width=4.1in]{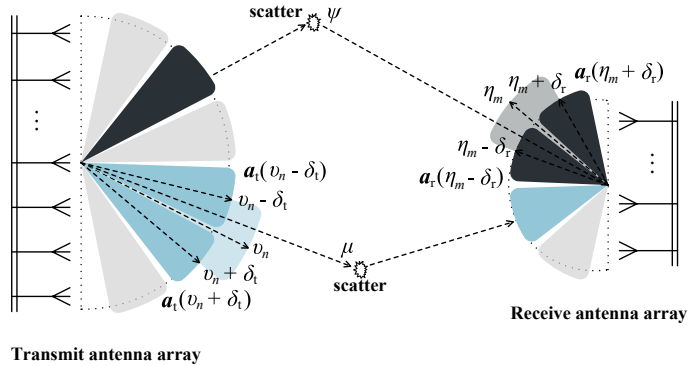}
\caption{An example of transmit and receive auxiliary beam pair design. The two beams in the transmit auxiliary beam pair steer towards $\nu_{n}-\delta_{\mathrm{t}}$ and $\nu_{n}+\delta_{\mathrm{t}}$. In the receive auxiliary beam pair, the two beams are steered towards $\eta_{m}-\delta_{\mathrm{r}}$ and $\eta_{m}+\delta_{\mathrm{r}}$.}
\end{figure}
In this section, the basic design principles for using an auxiliary beam pair to estimate the AoD and AoA in both single-path and multi-path channels are described.
\subsection{Single-path AoD/AoA estimation} The single-path channel is expressed as $\bm{H}_{1}=\alpha\bm{a}_{\mathrm{r}}(\phi)\bm{a}_{\mathrm{t}}^{*}(\theta)$, where $\alpha=\sqrt{N_{\mathrm{tot}}M_{\mathrm{tot}}}g$ and the path index is dropped for simplicity. Denote by $\mu=\frac{2\pi}{\lambda}d_{\mathrm{t}}\sin(\theta)$ and $\psi=\frac{2\pi}{\lambda}d_{\mathrm{r}}\sin(\phi)$ the transmit and receive spatial frequencies. The array response vectors for the transmitter and receiver can therefore be rewritten as $\bm{a}_{\mathrm{t}}(\mu)=\frac{1}{\sqrt{N_{\mathrm{tot}}}}\left[1, e^{j\mu},\cdots, e^{j\left(N_{\mathrm{tot}}-1\right)\mu} \right]^{\mathrm{T}}$ and $\bm{a}_{\mathrm{r}}(\psi)=\frac{1}{\sqrt{M_{\mathrm{tot}}}}\left[1, e^{j\psi},\cdots, e^{j\left(M_{\mathrm{tot}}-1\right)\psi} \right]^{\mathrm{T}}$.

Denote by $N_{\mathrm{K}}$ and $M_{\mathrm{K}}$ the total number of auxiliary beam pairs probed from the transmitter and receiver. As shown in Fig.~2, each auxiliary beam pair contains two consecutive analog beams in the angular domain. The total numbers of analog transmit and receive beams are therefore $N_{\mathrm{K}}+1$ and $M_{\mathrm{K}}+1$. That is, every two consecutive auxiliary beam pairs share one common analog beam. Consider the $n$-th ($n=1,\cdots,N_{\mathrm{K}}$) auxiliary beam pair formed by the transmitter as $\bm{a}_{\mathrm{t}}(\nu_{n}-\delta_{\mathrm{t}})$ and $\bm{a}_{\mathrm{t}}(\nu_{n}+\delta_{\mathrm{t}})$, where $\nu_{n}$ is the boresight angle of the $n$-th transmit auxiliary beam pair, and $\delta_{\mathrm{t}}$ represents half of the main beamforming region for the transmitter to ensure seamless coverage. Similarly, consider the $m$-th ($m=1,\cdots,M_{\mathrm{K}}$) auxiliary beam pair formed by the receiver as $\bm{a}_{\mathrm{r}}(\eta_{m}-\delta_{\mathrm{r}})$ and $\bm{a}_{\mathrm{r}}(\eta_{m}+\delta_{\mathrm{r}})$, where $\eta_{m}$ and $\delta_{\mathrm{r}}$ are similarly defined to $\nu_{n}$ and $\delta_{\mathrm{t}}$. Denote by $\mathcal{A}^{\mathrm{t}}_{n}=\left[\nu_{n}-\delta_{\mathrm{t}},\nu_{n}+\delta_{\mathrm{t}}\right]$ and $\mathcal{A}^{\mathrm{r}}_{m}=\left[\eta_{m}-\delta_{\mathrm{r}},\eta_{m}+\delta_{\mathrm{r}}\right]$ the main probing ranges of the $n$-th and $m$-th transmit and receive auxiliary beam pairs. We further assume that the main probing ranges of auxiliary beam pairs are disjoint, i.e., $\cap_{n=1}^{N_{\mathrm{K}}}\mathcal{A}^{\mathrm{t}}_{n}=\emptyset$ and $\cap_{m=1}^{M_{\mathrm{K}}}\mathcal{A}^{\mathrm{r}}_{m}=\emptyset$. This is reasonable if the number of deployed antennas is large, and the sidelobes are small. For given angular ranges $\Omega_{\mathrm{t}}$ and $\Omega_{\mathrm{r}}$ for the transmitter and receiver (e.g., $\Omega_{\mathrm{t}}=\left[-\pi/3,\pi/3\right]$ assuming a sectorized cellular structure and $\Omega_{\mathrm{r}}=\left[-\pi/2,\pi/2\right]$), $\cup_{n=1}^{N_{\mathrm{K}}}\mathcal{A}^{\mathrm{t}}_{n}=\Omega_{\mathrm{t}}$ and $\cup_{m=1}^{M_{\mathrm{K}}}\mathcal{A}^{\mathrm{r}}_{m}=\Omega_{\mathrm{r}}$. We assume that the transmit and receive beams are probed in a time division multiplexing (TDM) manner. For a given analog receive beam, all analog transmit beams are successively probed by the transmitter. This process continues until all $M_{\mathrm{K}}+1$ analog receive beams have been probed.

To estimate the transmit spatial frequency $\mu$, for a given analog receive beam, say, $\bm{a}_{\mathrm{r}}(\eta_{m}+\delta_{\mathrm{r}})$ and $\bm{a}_{\mathrm{t}}(\nu_{n}-\delta_{\mathrm{t}})$ in the $n$-th transmit auxiliary beam pair, the received signal is derived as
\begin{eqnarray}\label{aoae}
y^{\Delta}_{n,m} =\alpha\bm{a}^{*}_{\mathrm{r}}(\eta_{m}+\delta_{\mathrm{r}})\bm{a}_{\mathrm{r}}(\psi)\bm{a}^{*}_{\mathrm{t}}(\mu)\bm{a}_{\mathrm{t}}(\nu_{n}-\delta_{\mathrm{t}})x_{1}+\bm{a}^{*}_{\mathrm{r}}(\eta_{m}+\delta_{\mathrm{r}})\bm{n}.
\end{eqnarray}
Assume that $\mu$ is within the half-power beamwidth of $\bm{a}_{\mathrm{t}}(\nu_{n}-\delta_{\mathrm{t}})$. The corresponding received signal strength can therefore be calculated as (assuming $|x_{1}|^{2}=1$ because of the single RF assumption)
\begin{align}
\chi_{n,m}^{\Delta}=&\left(y^{\Delta}_{n,m}\right)^{*}y^{\Delta}_{n,m}\label{lb}\\
\leq&\underbrace{|\alpha|^{2}\bm{a}_{\mathrm{t}}^{*}(\nu_{n}-\delta_{\mathrm{t}})\bm{a}_{\mathrm{t}}(\mu)\bm{a}^{*}_{\mathrm{t}}(\mu)\bm{a}_{\mathrm{t}}(\nu_{n}-\delta_{\mathrm{t}})}_{I_1}\label{ineq}\\
&+\underbrace{\bm{n}^{*}\bm{a}_{\mathrm{r}}(\eta_{m}+\delta_{\mathrm{r}})\bm{a}^{*}_{\mathrm{r}}(\eta_{m}+\delta_{\mathrm{r}})\bm{n}}_{I_2}\nonumber\\
&+\underbrace{\alpha\bm{a}_{\mathrm{t}}^{*}(\nu_{n}-\delta_{\mathrm{t}})\bm{a}_{\mathrm{t}}(\mu)\bm{a}^{*}_{\mathrm{r}}(\eta_{m}+\delta_{\mathrm{r}})\bm{n}}_{I_3}\nonumber\\
&+\underbrace{\alpha\bm{n}^{*}\bm{a}_{\mathrm{r}}(\eta_{m}+\delta_{\mathrm{r}})\bm{a}^{*}_{\mathrm{t}}(\mu)\bm{a}_{\mathrm{t}}(\nu_{n}-\delta_{\mathrm{t}})}_{I_4}\nonumber\\
\approx&|\alpha|^{2}\bm{a}_{\mathrm{t}}^{*}(\nu_{n}-\delta_{\mathrm{t}})\bm{a}_{\mathrm{t}}(\mu)\bm{a}^{*}_{\mathrm{t}}(\mu)\bm{a}_{\mathrm{t}}(\nu_{n}-\delta_{\mathrm{t}}).\label{app}
\end{align}%
The equality in (\ref{ineq}) is achieved if $\bm{a}_{\mathrm{r}}(\eta_{m}+\delta_{\mathrm{r}})=\bm{a}_{\mathrm{r}}(\psi)$. To approach $\bm{a}_{\mathrm{r}}(\eta_{m}+\delta_{\mathrm{r}})=\bm{a}_{\mathrm{r}}(\psi)$, i.e., to minimize the gap with the upper bound in (\ref{ineq}), the receiver calculates the received signal strength for every combination between analog transmit and receive beams. The analog receive beam that yields the highest received signal strength is then used in (\ref{aoae}). Because $\mu$ is within the half-power beamwidth of $\bm{a}_{\mathrm{t}}(\nu_n-\delta_{\mathrm{t}})$, by assuming large $N_{\mathrm{tot}}M_{\mathrm{tot}}$, $|I_1|\gg|I_2|,|I_3|,|I_4|$, which results in the approximation in (\ref{app}). In this paper, we employ (\ref{app}) to derive the following results and denote by $\chi_{n}^{\Delta}=|\alpha|^{2}\bm{a}_{\mathrm{t}}^{*}(\nu_{n}-\delta_{\mathrm{t}})\bm{a}_{\mathrm{t}}(\mu)\bm{a}^{*}_{\mathrm{t}}(\mu)\bm{a}_{\mathrm{t}}(\nu_{n}-\delta_{\mathrm{t}})$.

Similarly, using $\bm{a}_{\mathrm{t}}(\nu_{n}+\delta_{\mathrm{t}})$ in the $n$-th transmit auxiliary beam pair, the received signal after combining with $\bm{a}_{\mathrm{r}}(\eta_{m}+\delta_{\mathrm{r}})$ is
\begin{eqnarray}\label{aoasigma}
y^{\Sigma}_{n,m} =\alpha\bm{a}^{*}_{\mathrm{r}}(\eta_{m}+\delta_{\mathrm{r}})\bm{a}_{\mathrm{r}}(\psi)\bm{a}^{*}_{\mathrm{t}}(\mu)\bm{a}_{\mathrm{t}}(\nu_{n}+\delta_{\mathrm{t}})x_{1}+\bm{a}^{*}_{\mathrm{r}}(\eta_{m}+\delta_{\mathrm{r}})\bm{n}.
\end{eqnarray}
The corresponding received signal strength can be calculated as $\chi_{n,m}^{\Sigma}=\left(y^{\Sigma}_{n,m}\right)^{*}y^{\Sigma}_{n,m}$. Similar to (\ref{app}) and the definition of $\chi_{n}^{\Delta}$, by assuming that $\mu$ is within the half-power beamwidth of $\bm{a}_{\mathrm{t}}(\nu_{n}+\delta_{\mathrm{t}})$, we can obtain $\chi_{n}^{\Sigma}=|\alpha|^{2}\bm{a}_{\mathrm{t}}^{*}(\nu_{n}+\delta_{\mathrm{t}})\bm{a}_{\mathrm{t}}(\mu)\bm{a}^{*}_{\mathrm{t}}(\mu)\bm{a}_{\mathrm{t}}(\nu_{n}+\delta_{\mathrm{t}})$. Further,
\begin{eqnarray}\label{expa}
\chi_{n}^{\Delta}=|\alpha|^{2}\frac{\sin^{2}\left(\frac{N_{\mathrm{tot}}(\mu-\nu_{n})}{2}\right)}{\sin^{2}\left(\frac{\mu-\nu_{n}+\delta_{\mathrm{t}}}{2}\right)},\hspace{3mm} \chi_{n}^{\Sigma}=|\alpha|^{2}\frac{\sin^{2}\left(\frac{N_{\mathrm{tot}}(\mu-\nu_{n})}{2}\right)}{\sin^{2}\left(\frac{\mu-\nu_{n}-\delta_{\mathrm{t}}}{2}\right)},
\end{eqnarray}
where (\ref{expa}) is obtained via $\left|\sum_{\bar{m}=1}^{M}e^{-j(\bar{m}-1)\bar{x}}\right|^{2}=\frac{\sin^{2}\left(\frac{M\bar{x}}{2}\right)}{\sin^{2}\left(\frac{\bar{x}}{2}\right)}$. The ratio metric $\zeta_{n}^{\mathrm{AoD}}$ is defined as
\begin{eqnarray}
\zeta_{n}^{\mathrm{AoD}}=\frac{\chi_{n}^{\Delta}-\chi_{n}^{\Sigma}}{\chi_{n}^{\Delta}+\chi_{n}^{\Sigma}}=\frac{\sin^{2}\left(\frac{\mu-\nu_{n}-\delta_{\mathrm{t}}}{2}\right)-\sin^{2}\left(\frac{\mu-\nu_{n}+\delta_{\mathrm{t}}}{2}\right)}{\sin^{2}\left(\frac{\mu-\nu_{n}-\delta_{\mathrm{t}}}{2}\right)+\sin^{2}\left(\frac{\mu-\nu_{n}+\delta_{\mathrm{t}}}{2}\right)}=-\frac{\sin\left(\mu-\nu_{n}\right)\sin(\delta_{\mathrm{t}})}{1-\cos\left(\mu-\nu_{n}\right)\cos(\delta_{\mathrm{t}})}.\label{ori}
\end{eqnarray}

It can be seen from the last equality in (\ref{ori}) that $\zeta_{n}^{\mathrm{AoD}}\in[-1,1]$. As the beam pattern designed in this paper is different from that in the monopulse radar systems, the following lemma is presented to illustrate the monotonicity of the ratio metric for a given interval.
\begin{lemma1}
If $|\mu-\nu_{n}|<\delta_{\mathrm{t}}$, i.e., the transmit spatial frequency $\mu$ is within the range of $\left(\nu_{n}-\delta_{\mathrm{t}},\nu_{n}+\delta_{\mathrm{t}}\right)$, $\zeta_{n}^{\mathrm{AoD}}$ is a monotonically decreasing function of $\mu-\nu_{n}$ and invertible with respect to $\mu-\nu_{n}$.
\end{lemma1}
\begin{proof}
Denote by $z=\mu-\nu_{n}$,
\begin{equation}\label{diffmono}
\frac{d}{dz}\frac{\sin(z)\sin(\delta_{\mathrm{t}})}{\cos(z)\cos(\delta_{\mathrm{t}})-1}=\frac{\sin(\delta_{\mathrm{t}})\left[\cos(\delta_{\mathrm{t}})-\cos(z)\right]}{\left[\cos(z)\cos(\delta_{\mathrm{t}})-1\right]^2}.
\end{equation}
According to the definition of $\delta_{\mathrm{t}}$, $\delta_{\mathrm{t}}\in\left[0,\pi\right]$, and $\sin(\delta_{\mathrm{t}})\geq0$. As $|\mu-\nu_{n}|<\delta_{\mathrm{t}}$, $\cos(\delta_{\mathrm{t}})-\cos(z)>0$ for every $z\in\left[-\pi,\pi\right]$. This leads to (\ref{diffmono}) being nonnegative for every $z$.
\end{proof}
By using the ratio metric $\zeta_{n}^{\mathrm{AoD}}$ and based on Lemma 1, the estimated value of $\mu$ via the inverse function can therefore be derived as
\begin{equation}\label{aoangle}
\hat{\mu}_{n}=\nu_{n}-\arcsin\left(\frac{\zeta_{n}^{\mathrm{AoD}}\sin(\delta_{\mathrm{t}})-\zeta_{n}^{\mathrm{AoD}}\sqrt{1-\left(\zeta_{n}^{\mathrm{AoD}}\right)^{2}}\sin(\delta_{\mathrm{t}})\cos(\delta_{\mathrm{t}})}{\sin^{2}(\delta_{\mathrm{t}})+\left(\zeta_{n}^{\mathrm{AoD}}\right)^{2}\cos^{2}(\delta_{\mathrm{t}})}\right).
\end{equation}
Note that if $\zeta_{n}^{\mathrm{AoD}}$ is perfect, i.e., not impaired by noise, the transmit spatial frequency can be perfectly recovered, i.e., $\mu=\hat{\mu}_{n}$. The corresponding array response vector for the transmitter can then be constructed as $\bm{a}_{\mathrm{t}}(\hat{\theta})$ with the estimated AoD $\hat{\theta}=\arcsin\left(\lambda\hat{\mu}_{n}/2\pi d_{\mathrm{t}}\right)$.

So far, the estimation of single-path AoD is illustrated using the ratio metric calculated from the $n$-th auxiliary beam pair as $|\mu-\nu_{n}|<\delta_{\mathrm{t}}$ is assumed. In practice, however, it is not possible to know which auxiliary beam pair covers the AoD \emph{a prior} at either the transmitter or receiver. It is therefore necessary to form multiple auxiliary beam pairs to cover a given angular range and determine a performance metric that helps the receiver to identify the transmit auxiliary beam pair whose main probing range most likely covers the AoD. For all transmit auxiliary beam pairs with a given analog combining vector, a set of ratio metrics $\left\{\zeta_{1}^{\mathrm{AoD}},\cdots,\zeta_{n}^{\mathrm{AoD}},\cdots,\zeta_{N_{\mathrm{K}}}^{\mathrm{AoD}}\right\}$ are determined according to (\ref{ori}). If the receiver does not have any knowledge of $\nu_{n}$'s and $\delta_{\mathrm{t}}$, it can select the ratio metric that characterizes the AoD the best using the following lemma.
\begin{lemma2}
If $|\mu-\nu_{n}|<\delta_{\mathrm{t}}$, i.e., the transmit spatial frequency $\mu$ is within the main probing range of the $n$-th auxiliary beam pair, and $|\mu-(\nu_n-\delta_{\mathrm{t}})|\leq|\mu-(\nu_n+\delta_{\mathrm{t}})|$,
\begin{equation}
\chi_{n}^{\Delta}=\underset{n'=1,\cdots,N_{\mathrm{K}}}{\mathrm{max}} ~\left\{\chi_{n'}^{\Delta},\chi_{n'}^{\Sigma}\right\},\label{pp2}
\end{equation}
for a given analog receive combining vector assuming no noise. Similarly, if $|\mu-\nu_{n}|<\delta_{\mathrm{t}}$ and $|\mu-(\nu_n+\delta_{\mathrm{t}})|\leq|\mu-(\nu_n-\delta_{\mathrm{t}})|$,
\begin{equation}
\chi_{n}^{\Sigma}=\underset{n'=1,\cdots,N_{\mathrm{K}}}{\mathrm{max}} ~\left\{\chi_{n'}^{\Delta},\chi_{n'}^{\Sigma}\right\},\label{pp3}
\end{equation}
for a given analog receive combining vector assuming no noise.
\end{lemma2}
\begin{proof}
Assume $|\mu-(\nu_n-\delta_{\mathrm{t}})|\leq|\mu-(\nu_n+\delta_{\mathrm{t}})|$. As $|\mu-\nu_{n}|<\delta_{\mathrm{t}}$, $\mu$ is within the half-power beamwidth of $\bm{a}_{\mathrm{t}}(\nu_n-\delta_{\mathrm{t}})$ because $\delta_{\mathrm{t}}$ is set as half of the half-power beamwidth for the given antenna array. Denote by
\begin{equation}
\chi_{\mathrm{max}}=|\alpha|^2\left|\bm{a}^{*}_{\mathrm{t}}(\mu)\bm{a}_{\mathrm{t}}(\mu)\right|^2,
\end{equation}
we have $\chi_{n}^{\Delta}\in\left[\frac{1}{2}\chi_{\mathrm{max}},\chi_{\mathrm{max}}\right]$ and $\chi_{n}^{\Sigma}\in\left[0,\frac{1}{2}\chi_{\mathrm{max}}\right]$. Note that $\chi_{n}^{\Delta}=\chi_{n}^{\Sigma}=\frac{1}{2}\chi_{\mathrm{max}}$ only occurs when $\mu=\nu_n$. According to the design principle of the auxiliary beam pairs, the main probing ranges of auxiliary beam pairs are disjoint. We therefore have $\chi_{n'}^{\Delta}, \chi_{n'}^{\Sigma}\in[0,\frac{1}{2}\chi_{\mathrm{max}})$ for $n'=1,\cdots,N_{\mathrm{K}}$, $n'\neq n$. Hence, $\chi_{n}^{\Delta}=\underset{n'=1,\cdots,N_{\mathrm{K}}}{\mathrm{max}} ~\left\{\chi_{n'}^{\Delta},\chi_{n'}^{\Sigma}\right\}$. For $|\mu-(\nu_n+\delta_{\mathrm{t}})|\leq|\mu-(\nu_n-\delta_{\mathrm{t}})|$, (\ref{pp3}) is obtained in a similar fashion.
\end{proof}

Lemma 2 implies that if the beam with the highest received signal strength is selected, the probing range of the corresponding auxiliary beam pair covers the transmit spatial frequency to be estimated. To choose the paired beam with respect to the beam selected using Lemma 2, the received signal strengths of its two adjacent beams are tested. The adjacent beam with the highest received signal strength among the two is then selected.

To estimate the receive spatial frequency $\psi$, the ratio metric can be similarly computed as
\begin{eqnarray}\label{aodratio}
\zeta_{m}^{\mathrm{AoA}}=-\frac{\sin\left(\psi-\eta_{m}\right)\sin(\delta_{\mathrm{r}})}{1-\cos\left(\psi-\eta_{m}\right)\cos(\delta_{\mathrm{r}})}.
\end{eqnarray}
If $|\psi-\eta_{m}|<\delta_{\mathrm{r}}$, $\zeta_{m}^{\mathrm{AoA}}$ is invertible with respect to $\psi-\eta_{m}$, and the estimated value of $\psi$ via the inverse function can be obtained as
\begin{equation}\label{aodangle}
\hat{\psi}_{m}=\eta_{m}-\arcsin\left(\frac{\zeta_{m}^{\mathrm{AoA}}\sin(\delta_{\mathrm{r}})-\zeta_{m}^{\mathrm{AoA}}\sqrt{1-\left(\zeta_{m}^{\mathrm{AoA}}\right)^{2}}\sin(\delta_{\mathrm{r}})\cos(\delta_{\mathrm{r}})}{{\sin^{2}(\delta_{\mathrm{r}})+\left(\zeta_{m}^{\mathrm{AoA}}\right)^{2}\cos^{2}(\delta_{\mathrm{r}})}}\right).
\end{equation}
The corresponding receive array response vector can be constructed as $\bm{a}_{\mathrm{r}}(\hat{\phi})$ with the estimated AoA $\hat{\phi}=\arcsin\left(\lambda\hat{\psi}_{m}/2\pi d_{\mathrm{r}}\right)$. According to (\ref{aodratio}) and (\ref{aodangle}), for all receive auxiliary beam pairs with a given transmit beamforming vector, a set of ratio metrics $\left\{\zeta_{1}^{\mathrm{AoA}},\cdots,\zeta_{m}^{\mathrm{AoA}},\cdots,\zeta_{M_{\mathrm{K}}}^{\mathrm{AoA}}\right\}$ and a set of estimated receive spatial frequencies $\{$$\hat{\psi}_{1},$$\cdots,$$\hat{\psi}_{m},$$\cdots,$$\hat{\psi}_{M_{\mathrm{K}}}$$\}$ are obtained by the receiver. The receive spatial frequency, and therefore, the AoA estimated from the receive auxiliary beam pair determined using Lemma 2 is then selected.

If multiple paths exist in the propagation channel, the proposed algorithm would estimate the dominant path's AoD and AoA with the highest path gain with high probability. Consider $\bm{a}_{\mathrm{t}}(\nu_{n}-\delta_{\mathrm{t}})$ and $\bm{a}_{\mathrm{r}}(\eta_{m}+\delta_{\mathrm{r}})$, (\ref{aoae}) can be rewritten as
\begin{equation}
\acute{y}^{\Delta}_{n,m} =\sum_{\ell'=1}^{N_{\mathrm{p}}}\alpha_{\ell'}\bm{a}^{*}_{\mathrm{r}}(\eta_{m}+\delta_{\mathrm{r}})\bm{a}_{\mathrm{r}}(\psi_{\ell'})\bm{a}^{*}_{\mathrm{t}}(\mu_{\ell'})\bm{a}_{\mathrm{t}}(\nu_{n}-\delta_{\mathrm{t}})x_{1}+\bm{a}^{*}_{\mathrm{r}}(\eta_{m}+\delta_{\mathrm{r}})\bm{n},
\end{equation}
where $\alpha_{\ell'}=g_{\ell'}\sqrt{N_{\mathrm{tot}}M_{\mathrm{tot}}}$. In the absence of noise, the corresponding received signal strength is calculated as
\begin{align}
\acute{\chi}_{n,m}^{\Delta} =& \hspace{1mm} |\alpha_{\ell}|^{2}\left|\bm{a}^{*}_{\mathrm{t}}(\mu_{\ell})\bm{a}_{\mathrm{t}}(\nu_{n}+\delta_{\mathrm{t}})\right|^{2}\left|\bm{a}^{*}({\psi_{\ell}})\bm{a}_{\mathrm{r}}(\eta_{m}+\delta_{\mathrm{r}})\right|^{2}\nonumber\\
+&\hspace{1mm}\alpha_{\ell}\bm{a}^{*}({\psi_{\ell}})\bm{a}_{\mathrm{r}}(\eta_{m}+\delta_{\mathrm{r}})\bm{a}^{*}_{\mathrm{t}}(\mu_{\ell})\bm{a}_{\mathrm{t}}(\nu_{n}+\delta_{\mathrm{t}})\nonumber\\
&\times\sum_{\ell'=1,\ell'\neq \ell}^{N_{\mathrm{p}}}\alpha_{\ell'}\bm{a}^{*}_{\mathrm{t}}(\nu_{n}+\delta_{\mathrm{t}})\bm{a}_{\mathrm{t}}(\mu_{\ell'})\bm{a}_{\mathrm{r}}^{*}(\psi_{\ell'})\bm{a}_{\mathrm{r}}(\eta_{m}+\delta_{\mathrm{r}})\nonumber\\
+&\hspace{1mm}\alpha_{\ell}\bm{a}^{*}_{\mathrm{t}}(\nu_{n}+\delta_{\mathrm{t}})\bm{a}_{\mathrm{t}}(\mu_{\ell})\bm{a}_{\mathrm{r}}^{*}(\psi_{\ell})\bm{a}_{\mathrm{r}}(\eta_{m}+\delta_{\mathrm{r}})\nonumber\\
&\times\sum_{\ell'=1,\ell'\neq \ell}^{N_{\mathrm{p}}}\alpha_{\ell'}\bm{a}^{*}({\psi_{\ell'}})\bm{a}_{\mathrm{r}}(\eta_{m}+\delta_{\mathrm{r}})\bm{a}^{*}_{\mathrm{t}}(\mu_{\ell'})\bm{a}_{\mathrm{t}}(\nu_{n}+\delta_{\mathrm{t}})\nonumber\\
+&\hspace{1mm}\sum_{\ell'=1,\ell'\neq \ell}^{N_{\mathrm{p}}}|\alpha_{\ell'}|^{2}\left|\bm{a}^{*}_{\mathrm{t}}(\mu_{\ell'})\bm{a}_{\mathrm{t}}(\nu_{n}+\delta_{\mathrm{t}})\right|^{2}\left|\bm{a}^{*}({\psi_{\ell'}})\bm{a}_{\mathrm{r}}(\eta_{m}+\delta_{\mathrm{r}})\right|^{2}\label{mulas}\\
\overset{\textrm{a.s.}}{\underset{N_{\mathrm{tot}}M_{\mathrm{tot}}\rightarrow\infty}{\xrightarrow{\hspace*{1.5cm}}}}&\hspace{1mm}|\alpha_{\ell}|^{2}\left|\bm{a}^{*}_{\mathrm{t}}(\mu_{\ell})\bm{a}_{\mathrm{t}}(\nu_{n}+\delta_{\mathrm{t}})\right|^{2}\left|\bm{a}^{*}({\psi_{\ell}})\bm{a}_{\mathrm{r}}(\eta_{m}+\delta_{\mathrm{r}})\right|^{2}\label{asrelt}.
\end{align}
Assuming $\mu_{\ell}\in\left(\nu_{n}-\delta_{\mathrm{t}},\nu_{n}+\delta_{\mathrm{t}}\right)$ and $\psi_{\ell}\in\left(\eta_{m}-\delta_{\mathrm{r}},\eta_{m}+\delta_{\mathrm{r}}\right)$, as mmWave channels generally exhibit sparse structure in the angular domain such that the number of multi-path components is limited with relatively small angular spreads \cite{mmwavesparse1, rap}, by jointly employing directional transmit beamforming and receive combining with $N_{\mathrm{tot}}M_{\mathrm{tot}}\rightarrow\infty$, the last three terms, in particular, the sum terms in (\ref{mulas}) converge to zeros, and (\ref{asrelt}) is achieved. Similar to (\ref{asrelt}), $\acute{\chi}_{n,m}^{\Sigma}$ can be obtained. Using the asymptotic results of $\acute{\chi}_{n,m}^{\Delta}$ and $\acute{\chi}_{n,m}^{\Sigma}$ to calculate (\ref{ori}), the same ratio metric can be obtained. If $\ell=\underset{\ell'=1,\cdots,N_{\mathrm{p}}}{\mathrm{argmax}} ~\alpha_{\ell'}$, the dominant path $\ell$ can be identified almost surely (a.s.) via simple power comparison \cite{omar1}. In the proposed method, a total number of $(N_{\mathrm{K}}+1)\times(M_{\mathrm{K}}+1)$ attempts are required by the receiver to simultaneously estimate the single-path AoD and AoA. Exploiting the proposed algorithm to estimate the multi-path components with multiple RF chains is described in Section III-E.
\subsection{Comparison between auxiliary beam pair design and monopulse radar waveform design}
\begin{figure}
\centering
\subfigure[]{%
\includegraphics[width=3.15in]{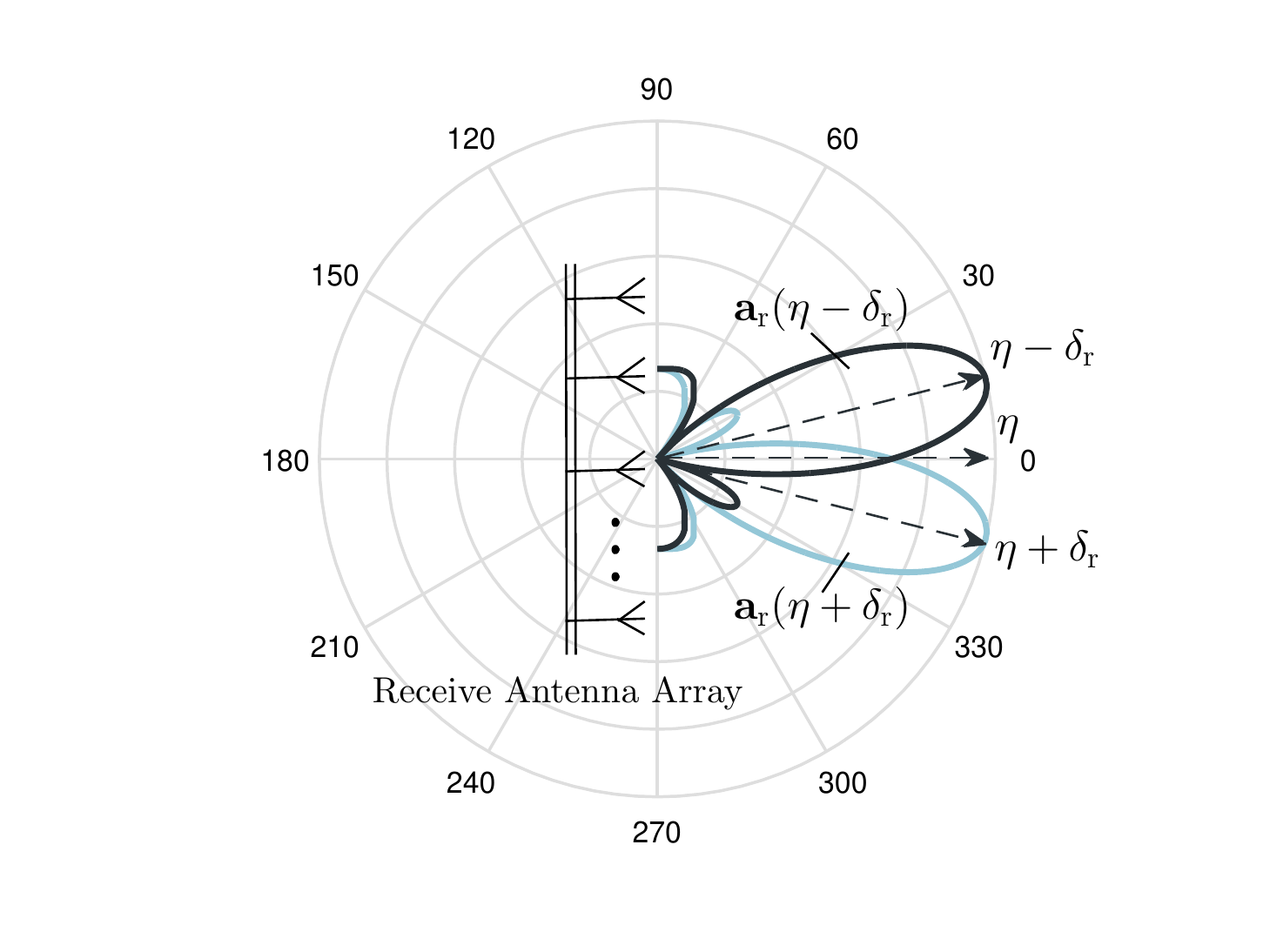}
\label{fig:subfigure1}}
\subfigure[]{%
\includegraphics[width=3.15in]{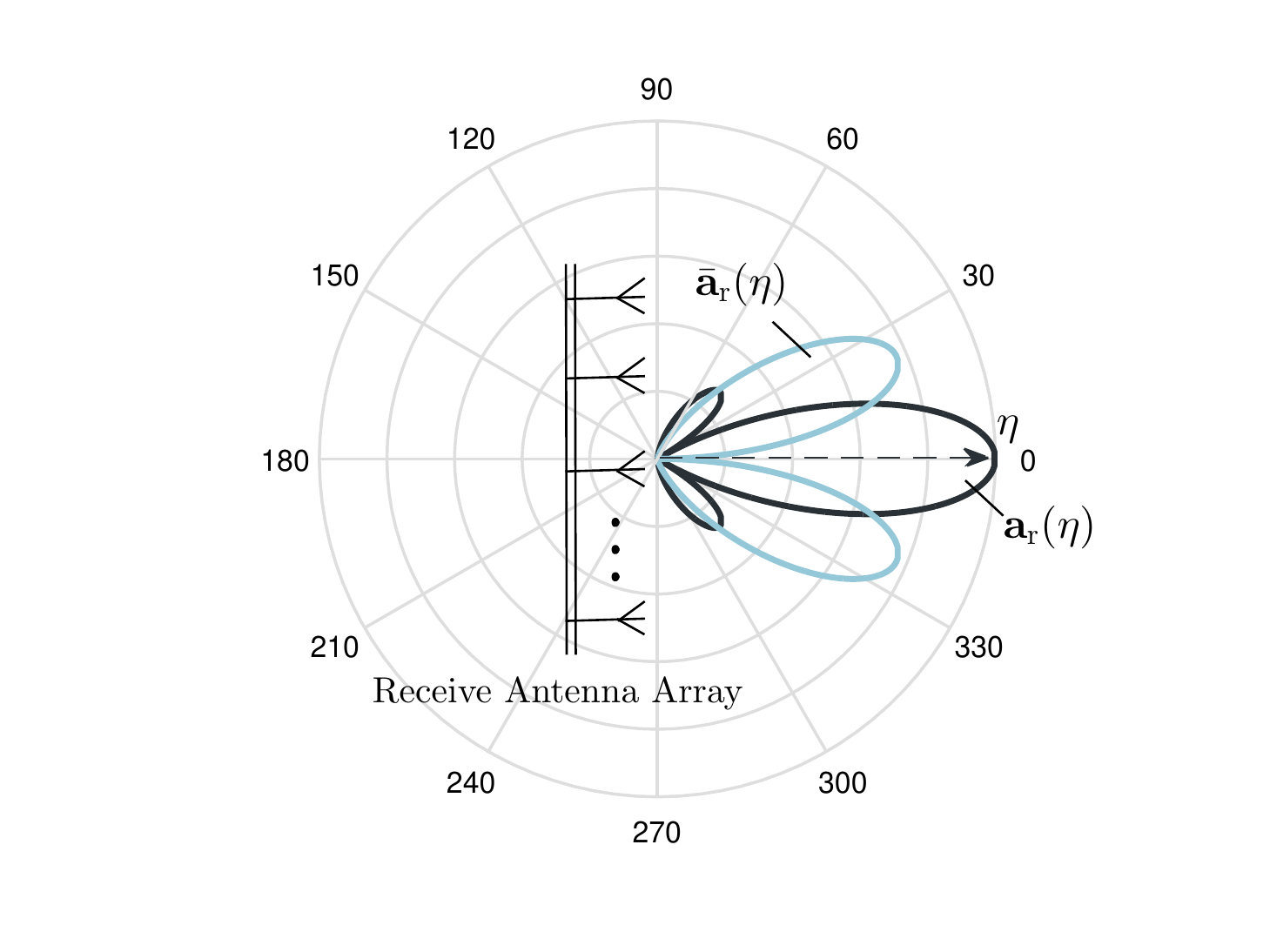}
\label{fig:subfigure2}}
\caption{(a) Beam pattern in the proposed auxiliary beam pair design. (b) Beam pattern in the conventional monopulse radar.}
\label{fig:figure}
\end{figure}
In this part, the key differences between the auxiliary beam pair based approach and monopulse radar systems are summarized in terms of the beam pattern design and compatibility with communications systems.
\subsubsection{Beam pattern}: The beam pattern in the proposed auxiliary beam pair exhibits the same form as the array response vector for linear arrays. Using the example shown in Fig.~3(a), the two receive combining vectors are constructed as $\bm{a}_{\mathrm{r}}(\eta-\delta_{\mathrm{r}})=\frac{1}{\sqrt{M_{\mathrm{tot}}}}\left[1,e^{j(\eta-\delta_{\mathrm{r}})},\cdots,e^{j(M_{\mathrm{tot}}-1)(\eta-\delta_{\mathrm{r}})}\right]^{\mathrm{T}}$ and $\bm{a}_{\mathrm{r}}(\eta+\delta_{\mathrm{r}})=\frac{1}{\sqrt{M_{\mathrm{tot}}}}\left[1,e^{j(\eta+\delta_{\mathrm{r}})},\cdots,e^{j(M_{\mathrm{tot}}-1)(\eta+\delta_{\mathrm{r}})}\right]^{\mathrm{T}}$. The beam pattern design for the monopulse radar is provided in Fig.~3(b). In monopulse radar, the beam pair comprises a sum beam and a difference beam, denoted by $\bm{a}_{\mathrm{r}}(\eta)$ and $\bar{\bm{a}}_{\mathrm{r}}(\eta)$. The sum beam $\bm{a}_{\mathrm{r}}(\eta)$ also exhibits the same structure as the array response vector for the linear array, and steers towards $\eta$. Hence,
\begin{equation}\label{summono1}
\bm{a}_{\mathrm{r}}(\eta)=\frac{1}{\sqrt{M_{\mathrm{tot}}}}\left[1,e^{j\eta},\cdots,e^{j(M_{\mathrm{tot}}-1)\eta}\right]^{\mathrm{T}}.
\end{equation}
The difference beam, however, exhibits a different structure from the array response vector for the linear array, and is constructed as
\begin{equation}\label{diffmono1}
\bar{\bm{a}}_{\mathrm{r}}\left(\eta\right)=\frac{1}{\sqrt{M_{\mathrm{tot}}}}\left[1,e^{j\eta},\cdots,e^{j\left(M_{\mathrm{tot}}/2-1\right)\eta},-e^{j\frac{M_{\mathrm{tot}}}{2}\eta},\cdots,-e^{j\left(M_{\mathrm{tot}}-1\right)\eta}\right]^{\mathrm{T}}. \end{equation}
As can be seen from (\ref{diffmono1}) and Fig.~3(b), the difference beam $\bar{\bm{a}}_{\mathrm{r}}\left(\eta\right)$ steers a null towards the boresight of the corresponding sum beam $\bm{a}_{\mathrm{r}}\left(\eta\right)$. The two beams $\bar{\bm{a}}_{\mathrm{r}}\left(\eta\right)$ and $\bm{a}_{\mathrm{r}}\left(\eta\right)$ in Fig.~3(b) can actually be visualized as a single beam with two large sidelobes such that each sidelobe is half of the power of the main lobe. Hence, while the proposed auxiliary beam pair design can simply rely on the well-defined Fourier transform (DFT)-type beam codebooks \cite{singh2}, the monopulse radar approach needs new beam codebooks due to the special structure of the beam patterns of difference beams, which may require extensive implementation efforts.
\subsubsection{Estimation overhead}: In monopulse radar, due to the special beam pair structure, the angular coverage provided by a given beam pair is approximately the half-power beamwidth of the corresponding sum beam. This is because the difference beam is only used in assisting the sum beam to perform the angle estimation, not providing angular coverage. In the proposed auxiliary beam pair design, however, all auxiliary beams are used to cover a given angular range. For the same number of antennas, the number of required sum beams in the monopulse design is approximately the same as the total number of beams used in the auxiliary beam pair design. As each sum beam is associated with a distinct difference beam, the estimation overhead required by the monopulse beam pair design almost doubles that of the proposed approach.
\subsection{Quantization and feedback options}
In a closed-loop system, the receiver can either quantize the ratio metric that characterizes the AoD or the estimated transmit spatial frequency, which will lead to different levels of angular resolution. Denote by $\mathcal{U}$ and $\mathcal{V}$ the codebooks for quantizing the ratio metric and the estimated transmit spatial frequency. The design of $\mathcal{U}$ is intricate as the ratio metric is a non-linear operation on $\mu$, $\nu_{n}$'s and $\delta_{\mathrm{t}}$. One example showing the density distribution of the ratio metric is provided in Fig.~4(a). In this example, both the single-path's AoD and AoA are uniformly distributed within $\left[-90^{\circ},90^{\circ}\right]$, with $N_{\mathrm{tot}}=16$, $\delta_{\mathrm{t}}=\pi/32$ and $M_{\mathrm{tot}}=1$ in the absence of noise. A total number of $3\times10^5$ samples are collected to plot the density distribution. It is observed that the ratio metric is non-uniformly distributed within the interval of $[-1,1]$ and symmetric with respect to $0$. Quantizing the estimated transmit spatial frequency is straightforward, i.e., if the receivers are uniformly dropped, it is uniformly distributed in the range of interest (see Fig.~4(b) which uses the same setup as in Fig.~4(a)). The receiver, however, requires the knowledge of $\nu_{n}$'s and $\delta_{\mathrm{t}}$ to estimate the transmit spatial frequency, which is unknown to the receiver. As $\nu_{n}$'s and $\delta_{\mathrm{t}}$ are parameters to form analog transmit beamforming vectors, they can be periodically broadcasted from the transmitter.
\begin{figure}
\begin{center}
\subfigure[]{%
\includegraphics[width=2.9in]{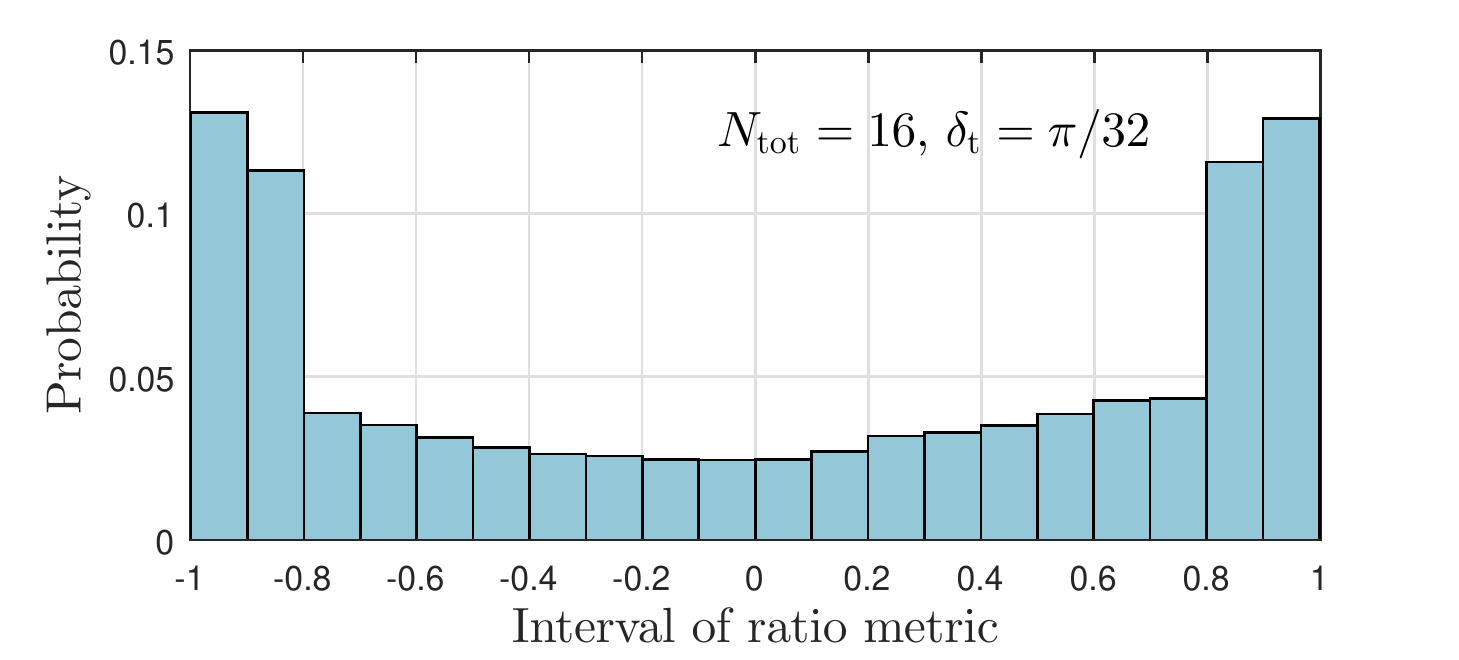}
\label{fig:subfigure1}}
\hspace{-3.5mm}
\subfigure[]{%
\includegraphics[width=2.9in]{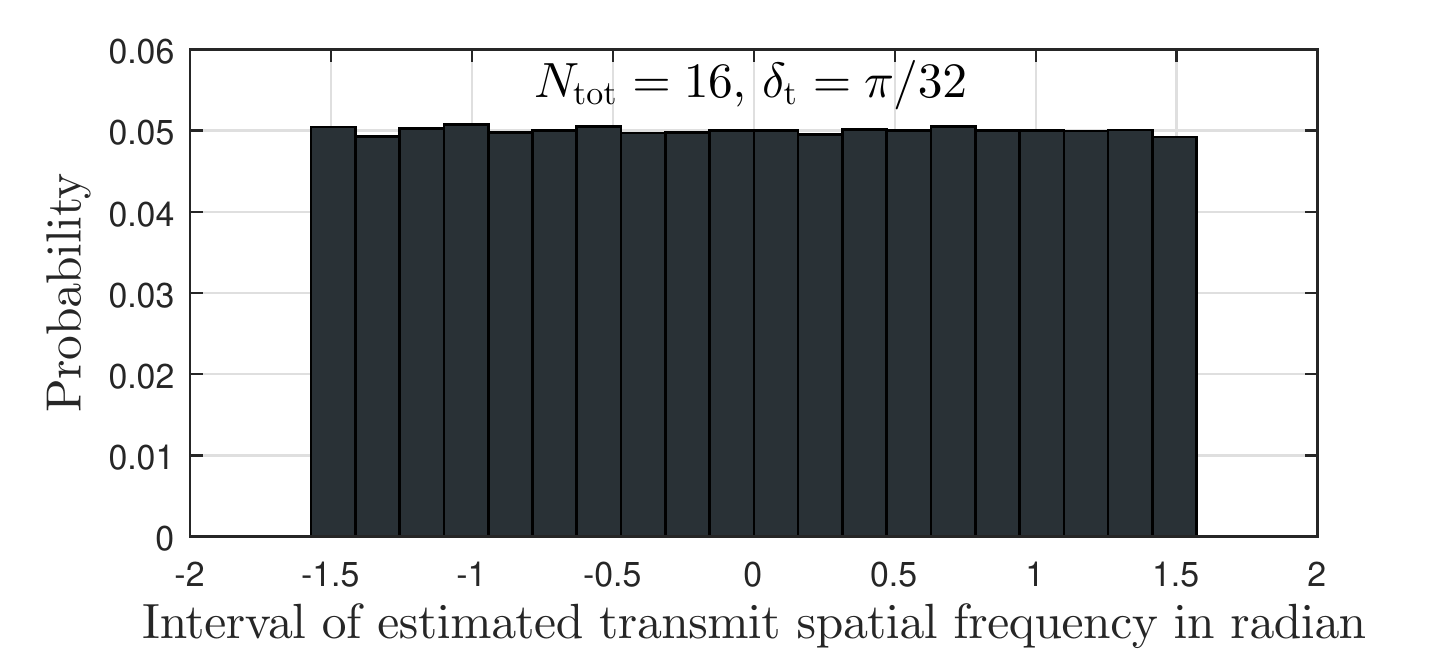}
\label{fig:subfigure2}}
\subfigure[]{%
\includegraphics[width=2.9in]{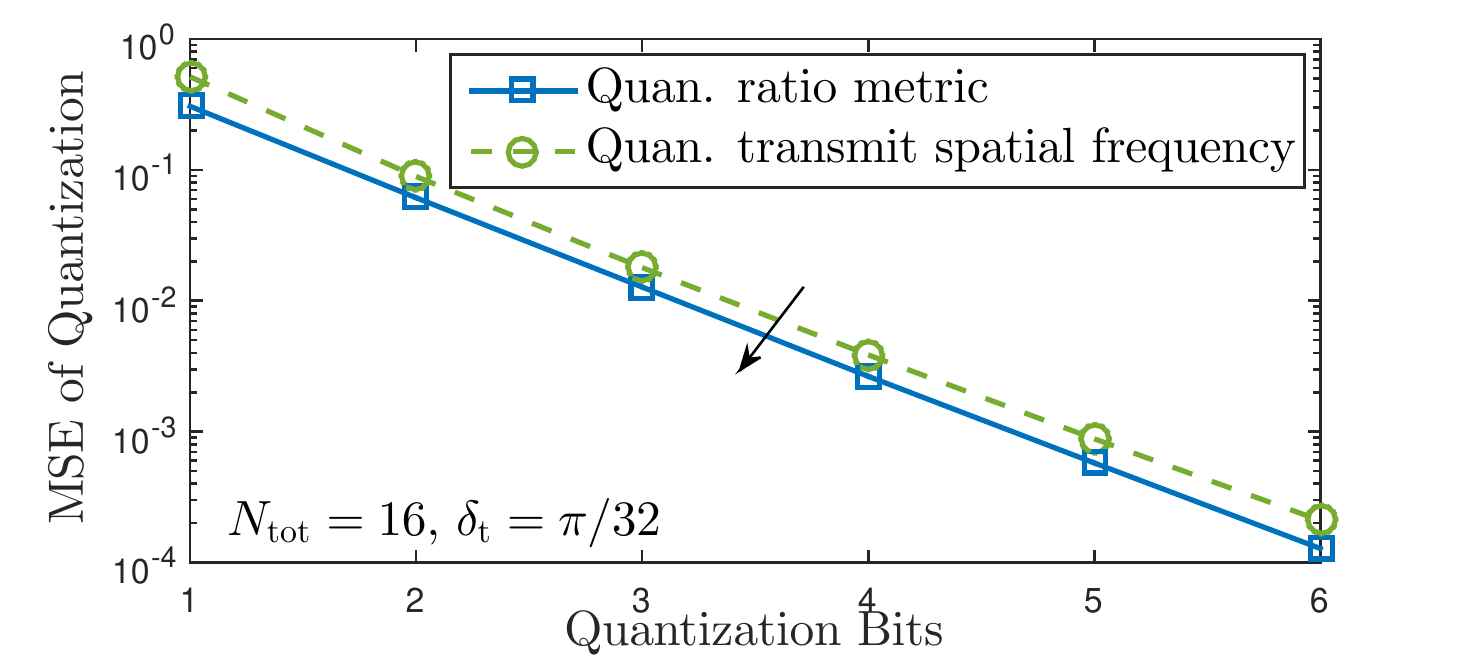}
\label{fig:subfigure3}}
\hspace{-3.5mm}
\subfigure[]{%
\includegraphics[width=3.2in]{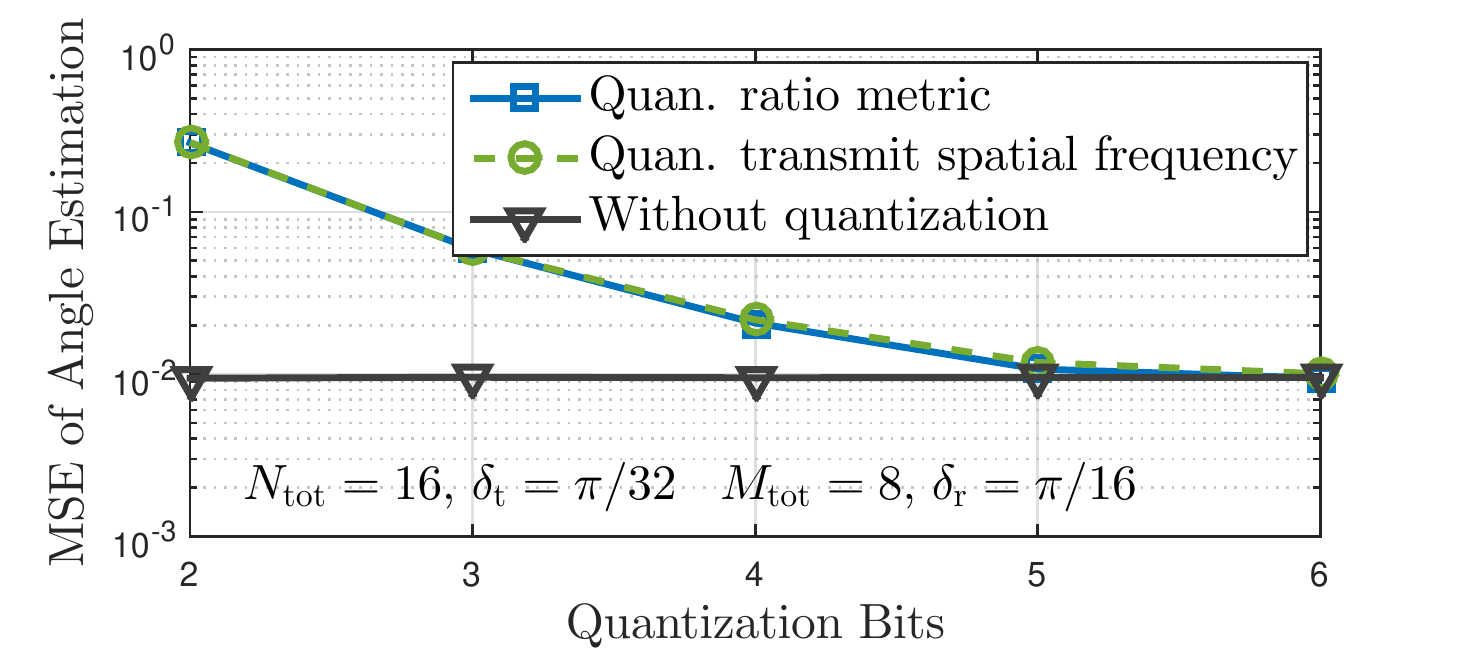}
\label{fig:subfigure3}}
\caption{(a) Probability density distribution of the ratio metric. (b) Probability density distribution of the estimated transmit spatial frequency. (c) Mean squared error performance of quantizing the ratio metric and estimated transmit spatial frequency. (d) Mean squared error performance of angle estimation with different numbers of quantization bits.}
\label{fig:figure}
\end{center}
\end{figure}

Though the codebook design for quantizing the ratio metric is relatively complicated, more design degrees of freedom can be offered for optimizing $\mathcal{U}$ than those for optimizing $\mathcal{V}$, of which the codewords are simply uniformly distributed in a given interval. For instance, the codebook $\mathcal{U}$ can be numerically optimized by allocating more codewords in densely distributed portions of the ratio metric, e.g., $[-1,0.8]$ and $[0.8,1]$ in the example shown in Fig.~4(a), to improve the quantization resolution.

The mean squared error (MSE) performance of quantizing the estimated ratio metric and transmit spatial frequency is provided in Fig.~4(c) applying the same setup as in Figs.~4(a) and 4(b). Here, the MSE of quantization is defined as $\mathbb{E}\left[\left|\upsilon_{\mathrm{est}}-\upsilon_{\mathrm{quan}}\right|^{2}\right]$, where $\upsilon_{\mathrm{est}}$ represents the estimated ratio metric or transmit spatial frequency, and $\upsilon_{\mathrm{quan}}$ is the quantized version of the corresponding estimated ratio metric or transmit spatial frequency. It is observed from Fig.~4(c) that quantizing the estimated ratio metric yields better quantization performance than quantizing the estimated transmit spatial frequency. This, however, does not mean that the ratio metric quantization gives better AoD estimation performance than the transmit spatial frequency quantization. This is because the quantized transmit spatial frequency can be directly applied by the transmitter but the quantized ratio metric has to be converted to the transmit spatial frequency first according to (\ref{aoangle}).

In Fig.~4(d), the MSE performance of angle estimation is examined with different quantization bits for both the ratio metric quantization and transmit spatial frequency quantization. The transmitter setup is the same as in Fig.~4(c) while the receiver now employs $M_{\mathrm{tot}}=8$ antennas with $\delta_{\mathrm{r}}=\pi/16$. The SNR is set to $-10$dB. For the ratio metric quantization, the quantized ratio metric is first converted to the transmit spatial frequency according to (\ref{aoangle}), and then transformed to the estimated AoD in radian. Regarding the transmit spatial frequency quantization, the quantized transmit spatial frequency is directly transformed to the estimated AoD in radian. From Fig.~4(d), it can be observed that the angle estimation performance between the two quantization approaches is similar for different numbers of quantization bits. Further, with relatively high quantization resolution (e.g., $6$ quantization bits), the angle estimation performance of the two quantization methods approaches that without quantization.

Since quantizing the transmit spatial frequency requires additional signaling support from the transmitter, we assume that the ratio measure is quantized and fed back to the transmitter.
\subsection{Performance analysis of single-path's/dominant path's angle estimation} In this subsection, we derive the variance of single-path's/dominant path's angle estimate using the proposed auxiliary beam pair design.

Assume that the transmit spatial frequency $\mu$ falls in the boresight of one of the beams in the $n$-th auxiliary beam pair, i.e., $\bm{a}_{\mathrm{t}}(\nu_{n}+\delta_{\mathrm{t}})=\bm{a}_{\mathrm{t}}(\mu)$, and the receive spatial frequency satisfies $|\psi-\eta_{m}|<\delta_{\mathrm{r}}$. To derive the following lemma that characterizes the receive spatial frequency estimation performance of the proposed approach, $\psi=0$ is assumed.
\begin{lemma3}
In a single-path channel, the variance of receive spatial frequency estimate, i.e., $\mathbb{E}\left[\hat{\psi}^2\right]$, using the proposed auxiliary beam pair design is approximated as
\begin{equation}\label{varerror}
\sigma_{\hat{\psi}}^{2}\approx\frac{(1-\cos(\delta_{\mathrm{r}}))^2\left|\Upsilon_{\Sigma}\right|}{2|\alpha|^2\gamma\sin^{2}(\delta_{\mathrm{r}})\left|\bm{a}^{*}_{\mathrm{r}}(\psi)\bm{\Lambda}_{\Delta}\bm{a}_{\mathrm{r}}(\psi)\right|}\left[1+\left(\zeta_{m}^{\mathrm{AoA}}\right)^2\right],
\end{equation}
where $\bm{\Lambda}_{\Delta}=\bm{a}_{\mathrm{r}}(\eta_{m}-\delta_{\mathrm{r}})\bm{a}^{*}_{\mathrm{r}}(\eta_{m}-\delta_{\mathrm{r}})-\bm{a}_{\mathrm{r}}(\eta_{m}+\delta_{\mathrm{r}})\bm{a}^{*}_{\mathrm{r}}(\eta_{m}+\delta_{\mathrm{r}})$ and $\Upsilon_{\Sigma}=\bm{a}^{*}_{\mathrm{r}}(\eta_{m}-\delta_{\mathrm{r}})\bm{a}_{\mathrm{r}}(\eta_{m}-\delta_{\mathrm{r}})+\bm{a}^{*}_{\mathrm{r}}(\eta_{m}+\delta_{\mathrm{r}})\bm{a}_{\mathrm{r}}(\eta_{m}+\delta_{\mathrm{r}})$.
\end{lemma3}
\begin{proof}
See Appendix.
\end{proof}

For a multi-path channel, assume that the dominant path's transmit spatial frequency $\mu_{\ell}$ ($\ell\in\left\{1,\cdots,N_{\mathrm{p}}\right\}$) is identical to the boresight of one of the beams in the $n$-th auxiliary beam pair, i.e., $\bm{a}_{\mathrm{t}}(\nu_{n}+\delta_{\mathrm{t}})=\bm{a}_{\mathrm{t}}(\mu_{\ell})$. The corresponding receive spatial frequency satisfies $|\psi_{\ell}-\eta_m|<\delta_{\mathrm{r}}$. Assuming $\psi_{\ell}=0$, the following corollary characterizes the dominant path's receive spatial frequency estimation performance.
\begin{corollary1}
For a given multi-path channel, the variance of dominant path's receive spatial frequency estimate, i.e., $\mathbb{E}\left[\hat{\psi}_{\ell}^2\right]$, using the proposed auxiliary beam pair design is approximated as
\begin{equation}\label{varerrormultipath}
\sigma_{\hat{\psi_{\ell}}}^{2}\approx\frac{(1-\cos(\delta_{\mathrm{r}}))^2\left(\sigma^2\left|\Upsilon_{\Sigma}\right|+\sum_{\ell'=1,\ell'\neq\ell}^{N_{\mathrm{p}}}\left|\bm{a}^{*}_{\mathrm{t}}(\nu_{n}+\delta_{\mathrm{t}})\bm{G}^{*}_{\ell'}\bm{\Lambda}_{\Sigma}\bm{G}_{\ell'}\bm{a}_{\mathrm{t}}(\nu_{n}+\delta_{\mathrm{t}})\right|\right)}{2|\alpha|^2\sin^{2}(\delta_{\mathrm{r}})\left|\bm{a}^{*}_{\mathrm{r}}(\psi_{\ell})\bm{\Lambda}_{\Delta}\bm{a}_{\mathrm{r}}(\psi_{\ell})\right|}\left[1+\left(\zeta_{m,\ell}^{\mathrm{AoA}}\right)^2\right],
\end{equation}
where $\bm{G}_{\ell'}=\alpha_{\ell'}\bm{a}_{\mathrm{r}}(\psi_{\ell'})\bm{a}^{*}_{\mathrm{t}}(\mu_{\ell'})$, $\zeta_{m,\ell}^{\mathrm{AoA}}=-\frac{\sin\left(\psi_{\ell}-\eta_{m}\right)\sin(\delta_{\mathrm{r}})}{1-\cos\left(\psi_{\ell}-\eta_{m}\right)\cos(\delta_{\mathrm{r}})}$, and $\bm{\Lambda}_{\Sigma}=\bm{a}_{\mathrm{r}}(\eta_{m}-\delta_{\mathrm{r}})\bm{a}^{*}_{\mathrm{r}}(\eta_{m}-\delta_{\mathrm{r}})+\bm{a}_{\mathrm{r}}(\eta_{m}+\delta_{\mathrm{r}})\bm{a}^{*}_{\mathrm{r}}(\eta_{m}+\delta_{\mathrm{r}})$.
\end{corollary1}
The approximated result in Corollary 1 is obtained by treating $N_{\Sigma}$ in (\ref{sum}) as the multi-path interference plus noise power of the sum channel output, which is computed as
\begin{equation}\label{sumchsinr}
N_{\Sigma}=\sigma^2\left|\Upsilon_{\Sigma}\right|+\sum_{\ell'=1,\ell'\neq\ell}^{N_{\mathrm{p}}}\left|\bm{a}^{*}_{\mathrm{t}}(\nu_{n}+\delta_{\mathrm{t}})\bm{G}^{*}_{\ell'}\bm{\Lambda}_{\Sigma}\bm{G}_{\ell'}\bm{a}_{\mathrm{t}}(\nu_{n}+\delta_{\mathrm{t}})\right|.
\end{equation}

In Fig.~5(a), the numerical result of the variance of receive spatial frequency estimate using the auxiliary beam pair design is provided along with the analytical result given in (\ref{varerror}). A single-path channel is employed in the simulation with $M_{\mathrm{tot}}=4$, $N_{\mathrm{tot}}=8$ and $\delta_{\mathrm{r}}=\pi/8$. Additionally, we assume that $\psi=0$ and the steering angle of the transmit beam is identical to the AoD. One receive auxiliary beam pair is formed with zero boresight angle. It can be observed from Fig.~5(a) that for various SNR values, the gap between the analytically and numerically computed variances of angle estimate is marginal, which is mainly caused by the approximation in (\ref{varerror}).

In Fig.~5(b), the numerically and analytically calculated variances of dominant path's receive spatial frequency in the multi-path channel are plotted. Similar simulation assumptions are made to those in Fig.~5(a). In addition, a total number of $N_{\mathrm{p}}$ channel paths are assumed with equal path gain, and the corresponding AoDs and AoAs are uniformly distributed within $\left[-45^\circ,45^\circ\right]$. Further, we assume that the AoDs and AoAs of the interfering multi-paths have continuous values and are fixed for all drops. As can be seen from Fig.~5(b), with an increase in $N_{\mathrm{p}}$ from $1$ to $8$, the estimation performance of dominant path's receive spatial frequency significantly degrades. Similar observations can be obtained by reducing the number of receive antennas $M_{\mathrm{tot}}$ from $4$ to $2$. For various numbers of multi-paths and receive antennas, the performance gap between the numerically computed variance of dominant path's receive spatial frequency estimate and that calculated from (\ref{varerrormultipath}) is marginal, which validates the analysis.
\begin{figure}
\begin{center}
\subfigure[]{%
\includegraphics[width=2.9in]{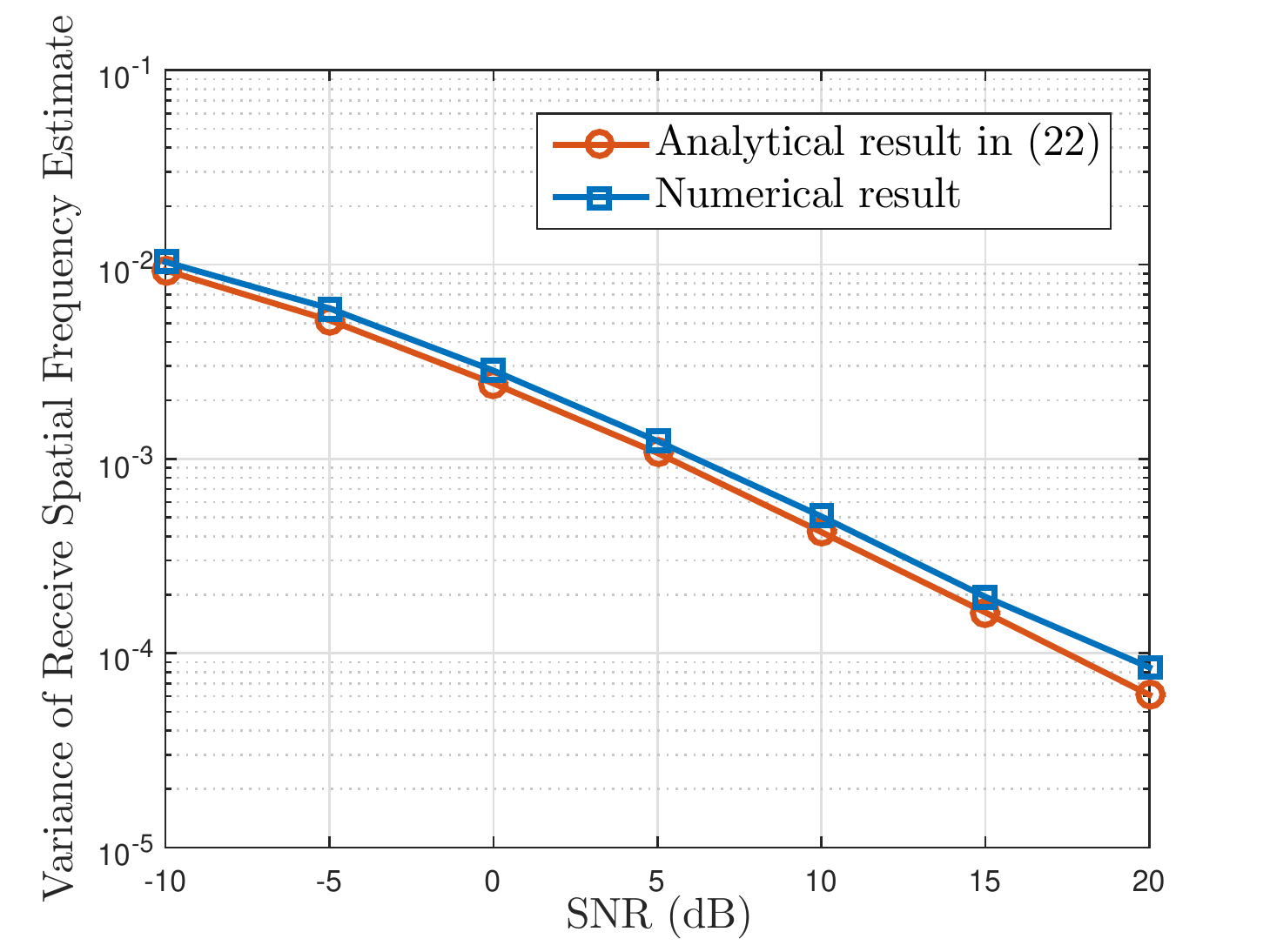}
\label{fig:subfigure1}}
\hspace{-3.5mm}
\subfigure[]{%
\includegraphics[width=2.9in]{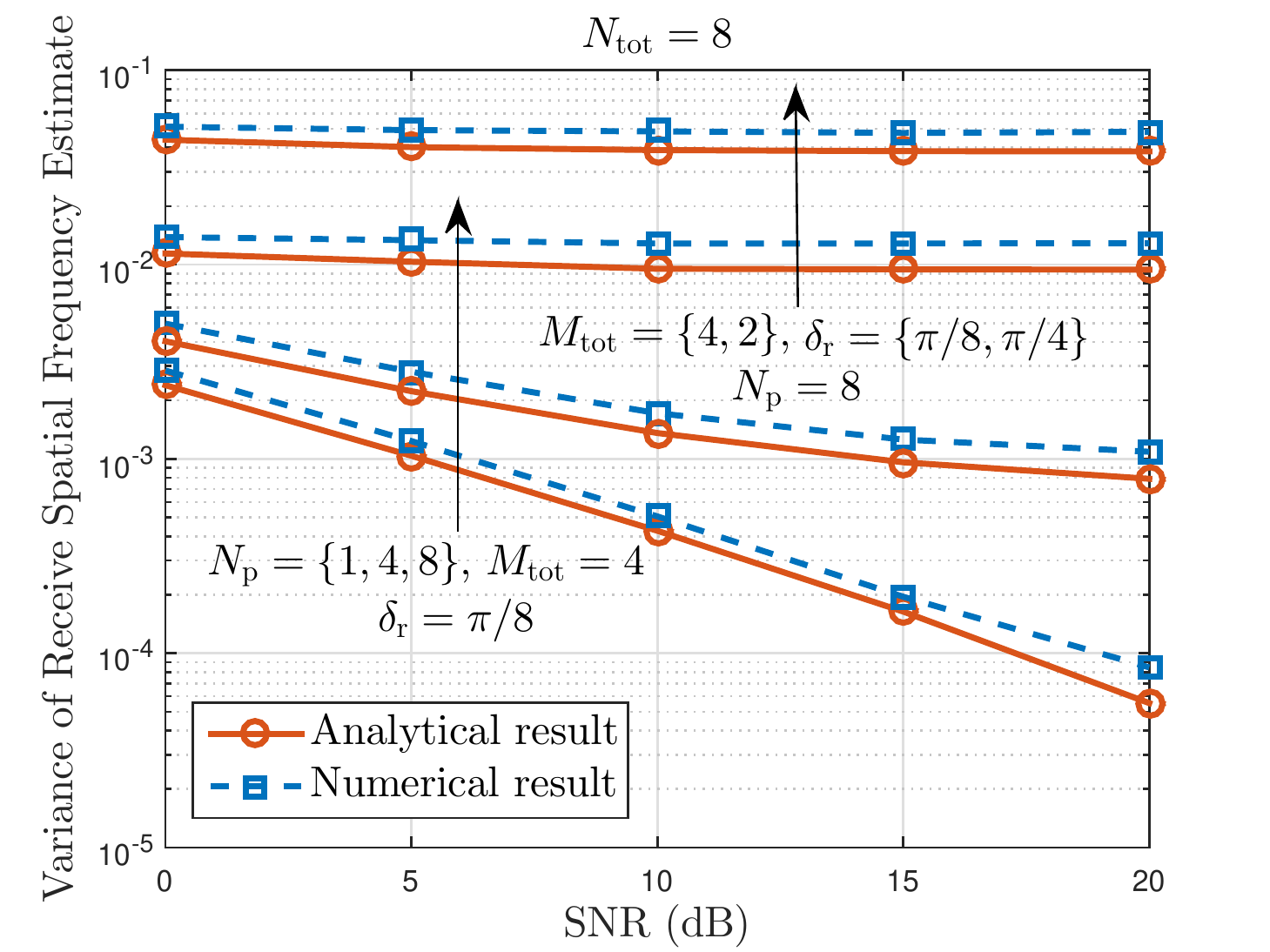}
\label{fig:subfigure2}}
\caption{(a) Analytical and numerical results of variance of single-path's receive spatial frequency estimate assuming $\psi=0$. (b) Analytical and numerical results of variance of dominant-path's receive spatial frequency estimate in a multi-path channel.}
\label{fig:figure}
\end{center}
\end{figure}
\subsection{Multi-path AoD/AoA estimation} If the transmitter and receiver have $N_{\mathrm{RF}}$ and $M_{\mathrm{RF}}$ RF chains, we propose that $N_{\mathrm{RF}}$ and $M_{\mathrm{RF}}$ analog transmit and receive beams are simultaneously probed for a given time instant. This is more efficient than forming one beam at a time using a single RF chain to cover a given angular range. By selecting the best transmit and receive auxiliary beam pairs from all beamforming and combining vectors, high-resolution multi-path AoD and AoA estimates can be obtained. To facilitate the selection of the best transmit and receive auxiliary beam pairs, the simultaneously formed analog transmit and receive beams are probed towards random directions. Pseudo-random sequence (e.g., m-sequence) scrambled with the beam specific identity (ID) is transmitted along with the analog transmit beam. By detecting the pseudo-random sequence, the receiver is able to differentiate the simultaneously probed analog transmit beams.

Denote by $N_{\mathrm{T}}$ and $M_{\mathrm{T}}$ the total numbers of probings performed by the transmitter and receiver. Denote by $\mathcal{F}_{\mathrm{T}}=\left\{\bm{a}_{\mathrm{t}}(\nu_{n}\pm\delta_{\mathrm{t}}), n=1,\cdots,N_{\mathrm{K}}\right\}$ and $\mathcal{W}_{\mathrm{T}}=\left\{\bm{a}_{\mathrm{r}}(\eta_{m}\pm\delta_{\mathrm{r}}), m=1,\cdots,M_{\mathrm{K}}\right\}$ the codebooks of analog transmit and receive steering vectors. The analog transmit and receive probing matrices are constructed by concatenating all successively probed analog transmit precoding and receive combining matrices. For instance, denote by $\bm{F}_{\mathrm{T}}$ and $\bm{W}_{\mathrm{T}}$ the analog transmit and receive probing matrices, we have $\bm{F}_{\mathrm{T}}=\left[\bm{F}_{1},\cdots,\bm{F}_{n_{\mathrm{t}}},\cdots,\bm{F}_{N_{\mathrm{T}}}\right]$ and $\bm{W}_{\mathrm{T}}=\left[\bm{W}_{1},\cdots,\bm{W}_{m_{\mathrm{t}}},\cdots,\bm{W}_{M_{\mathrm{T}}}\right]$, where $\bm{F}_{n_{\mathrm{t}}}\in\mathbb{C}^{N_{\mathrm{tot}}\times N_{\mathrm{RF}}}$ represents the $n_{\mathrm{t}}$-th probing formed by the transmitter, $\bm{W}_{m_{\mathrm{t}}}\in\mathbb{C}^{M_{\mathrm{tot}}\times M_{\mathrm{RF}}}$ is the $m_{\mathrm{t}}$-th probing at the receiver. Each column in $\bm{F}_{n_{\mathrm{t}}}$ and $\bm{W}_{m_{\mathrm{t}}}$ is randomly chosen from $\mathcal{F}_{\mathrm{T}}$ and $\mathcal{W}_{\mathrm{T}}$. The reason for simultaneously steering the transmit and receive beams towards random directions other than successive or predefined directions will be elaborated on later. The transmit and receive probings are conducted in a TDM manner. That is, for a given probing at the receiver, e.g., $\bm{W}_{m_{\mathrm{t}}}$, $N_{\mathrm{T}}$ consecutive probings $\bm{F}_{1},\cdots,\bm{F}_{N_{\mathrm{T}}}$ are performed at the transmitter. This process iterates until all $M_{\mathrm{T}}$ probings have been executed by the receiver. As each beam in $\mathcal{F}_{\mathrm{T}}$ is associated with a distinct beam ID, the receiver can identify a specific beam in $\bm{F}_{\mathrm{T}}$, and therefore, a specific transmit auxiliary beam pair.

Assume that $\mu_{\ell}\in\left(\nu_{n}-\delta_{\mathrm{t}},\nu_{n}+\delta_{\mathrm{t}}\right)$ and $\psi_{\ell}\in\left(\eta_{m}-\delta_{\mathrm{r}},\eta_{m}+\delta_{\mathrm{r}}\right)$ for a given $\ell\in\left\{1,\cdots,N_{\mathrm{p}}\right\}$ where $N_{\mathrm{p}}$ is the total number of paths. To estimate path-$\ell$'s transmit spatial frequency $\mu_{\ell}$, consider a given probing at the receiver, e.g., $\bm{W}_{m_{\mathrm{t}}}$, the resultant matrix by concatenating the $N_{\mathrm{T}}N_{\mathrm{RF}}$ transmit beamforming vectors in the absence of noise is obtained as
\begin{eqnarray}
\bm{Y}_{m_{\mathrm{t}}}&=&\bm{W}_{m_{\mathrm{t}}}^{*}\bm{H}\bm{F}_{\mathrm{T}}\bm{X},\label{sp}
\end{eqnarray}
where $\bm{X}$ is a diagonal matrix carrying $N_{\mathrm{T}}N_{\mathrm{RF}}$ training symbols on its diagonal. For simplicity, we set $\bm{X}=\bm{I}_{N_{\mathrm{T}}N_{\mathrm{RF}}}$. By detecting the beam ID, the receiver is able to locate a specific transmit auxiliary beam pair in $\bm{F}_{\mathrm{T}}$. For instance, $\left[\bm{F}_{\mathrm{T}}\right]_{:,u}=\bm{a}_{\mathrm{t}}(\nu_{n}-\delta_{\mathrm{t}})$, and $\left[\bm{F}_{\mathrm{T}}\right]_{:,v}=\bm{a}_{\mathrm{t}}(\nu_{n}+\delta_{\mathrm{t}})$, $u,v\in\left\{1,\cdots,N_{\mathrm{T}}N_{\mathrm{RF}}\right\}$. Consider path-$\ell$ and $\bm{a}_{\mathrm{t}}(\nu_{n}-\delta_{\mathrm{t}})$, we have
\begin{eqnarray}\label{rn}
\left|\left[\bm{Y}_{m_{\mathrm{t}}}\right]_{\ell,u}\right|^{2}&=&\chi_{n,m_{\mathrm{t}},\ell}^{\Delta}+\sum_{\ell'=1,\ell'\neq \ell}^{N_{\mathrm{p}}}|\alpha_{\ell'}|^{2}\bigg[\left[\bm{W}_{m_{\mathrm{t}}}\right]_{:,\ell}^{*}\bm{a}_{\mathrm{r}}(\psi_{\ell'})\bm{a}^{*}_{\mathrm{t}}(\mu_{\ell'})\bm{a}_{\mathrm{t}}(\nu_{n}-\delta_{\mathrm{t}})\nonumber\\
&\times&\bm{a}^{*}_{\mathrm{t}}(\nu_{n}-\delta_{\mathrm{t}})\bm{a}_{\mathrm{t}}(\mu_{\ell'})\bm{a}^{*}_{\mathrm{r}}(\psi_{\ell'})\left[\bm{W}_{m_{\mathrm{t}}}\right]_{:,\ell}\bigg],
\end{eqnarray}
where $\chi_{n,m_{\mathrm{t}},\ell}^{\Delta}=|\alpha_{\ell}|^{2}\left|\left[\bm{W}_{m_{\mathrm{t}}}\right]_{:,\ell}^{*}\bm{a}_{\mathrm{r}}(\psi_{\ell})\right|^{2}\left|\bm{a}_{\mathrm{t}}^{*}(\mu_{\ell})\bm{a}_{\mathrm{t}}(\nu_{n}-\delta_{\mathrm{t}})\right|^{2}$. Exploiting the sparse nature of mmWave channels, if $N_{\mathrm{tot}}M_{\mathrm{tot}}\rightarrow\infty$, the second term in (\ref{rn}) converges to zero \cite{omar1}, resulting in $\left|\left[\bm{Y}_{m_{\mathrm{t}}}\right]_{\ell,u}\right|^{2} \overset{\textrm{a.s.}}{\rightarrow} \chi_{n,m_{\mathrm{t}},\ell}^{\Delta}$. Similarly, with respect to $\bm{a}_{\mathrm{t}}(\nu_{n}+\delta_{\mathrm{t}})$, we have $\left|\left[\bm{Y}_{m_{\mathrm{t}}}\right]_{\ell,v}\right|^{2}\overset{\textrm{a.s.}}{\rightarrow}\chi_{n,m_{\mathrm{t}},\ell}^{\Sigma}=|\alpha_{\ell}|^{2}\left|\left[\bm{W}_{m_{\mathrm{t}}}\right]_{:,\ell}^{*}\bm{a}_{\mathrm{r}}(\psi_{\ell})\right|^{2}\left|\bm{a}_{\mathrm{t}}^{*}(\mu_{\ell})\bm{a}_{\mathrm{t}}(\nu_{n}+\delta_{\mathrm{t}})\right|^{2}$. Using the asymptotic results to estimate $\mu_{\ell}$, the ratio metric is calculated as
\begin{eqnarray}\label{intratio}
\zeta_{n,\ell}^{\mathrm{AoD}}&=&\frac{\left|\left[\bm{Y}_{m_{\mathrm{t}}}\right]_{\ell,u}\right|^{2}-\left|\left[\bm{Y}_{m_{\mathrm{t}}}\right]_{\ell,v}\right|^{2}}{\left|\left[\bm{Y}_{m_{\mathrm{t}}}\right]_{\ell,u}\right|^{2}+\left|\left[\bm{Y}_{m_{\mathrm{t}}}\right]_{\ell,v}\right|^{2}}=\frac{\chi_{n,m_{\mathrm{t}},\ell}^{\Delta}-\chi_{n,m_{\mathrm{t}},\ell}^{\Sigma}}{\chi_{n,m_{\mathrm{t}},\ell}^{\Delta}+\chi_{n,m_{\mathrm{t}},\ell}^{\Sigma}}\\
&=&\frac{\left|\bm{a}_{\mathrm{t}}^{*}(\mu_{\ell})\bm{a}_{\mathrm{t}}(\nu_{n}-\delta_{\mathrm{t}})\right|^{2}-\left|\bm{a}_{\mathrm{t}}^{*}(\mu_{\ell})\bm{a}_{\mathrm{t}}(\nu_{n}+\delta_{\mathrm{t}})\right|^{2}}{\left|\bm{a}_{\mathrm{t}}^{*}(\mu_{\ell})\bm{a}_{\mathrm{t}}(\nu_{n}-\delta_{\mathrm{t}})\right|^{2}+\left|\bm{a}_{\mathrm{t}}^{*}(\mu_{\ell})\bm{a}_{\mathrm{t}}(\nu_{n}+\delta_{\mathrm{t}})\right|^{2}}\label{l1}.
\end{eqnarray}
According to Lemma 2 and $\mu_{\ell}\in\left(\nu_{n}-\delta_{\mathrm{t}},\nu_{n}+\delta_{\mathrm{t}}\right)$, $\chi_{n,m_{\mathrm{t}},\ell}^{\Delta}$ and $\chi_{n,m_{\mathrm{t}},\ell}^{\Sigma}$ are determined as the auxiliary beam pair of interest. The quantized version of $\zeta_{n,\ell}^{\mathrm{AoD}}$ is fed back to the transmitter. Upon receiving the feedback information, the transmitter estimates $\mu_{\ell}$ via (similar to (\ref{aodangle}))
\begin{equation}\label{aodangleR}
\hat{\mu}_{n,\ell}=\nu_{n}-\arcsin\left(\frac{\zeta_{n,\ell}^{\mathrm{AoD}}\sin(\delta_{\mathrm{t}})-\zeta_{n,\ell}^{\mathrm{AoD}}\sqrt{1-\left(\zeta_{n,\ell}^{\mathrm{AoD}}\right)^{2}}\sin(\delta_{\mathrm{t}})\cos(\delta_{\mathrm{t}})}{{\sin^{2}(\delta_{\mathrm{t}})+\left(\zeta_{n,\ell}^{\mathrm{AoD}}\right)^{2}\cos^{2}(\delta_{\mathrm{t}})}}\right).
\end{equation}
The corresponding transmit array response vector can be constructed as $\bm{a}_{\mathrm{t}}(\hat{\theta}_{\ell})$ with $\hat{\theta}_{\ell}=\arcsin\left(\lambda\hat{\mu}_{n,\ell}/2\pi d_{\mathrm{t}}\right)$. The above process is conducted with respect to each path $\ell$, and finally, $\hat{\bm{A}}_{\mathrm{t}}=\left[\bm{a}_{\mathrm{t}}(\hat{\theta}_{1}),\cdots,\bm{a}_{\mathrm{t}}(\hat{\theta}_{N_{\mathrm{p}}})\right]$ is constructed accounting for all estimated AoDs.

To estimate path-$\ell$'s AoA $\psi_{\ell}$, a given probing at the transmitter $\bm{F}_{n_{\mathrm{t}}}$ is considered by concatenating all $M_{\mathrm{T}}$ receive probings in the absence of noise, i.e.,
\begin{equation}
\bm{Y}_{n_{\mathrm{t}}}=\bm{W}_{\mathrm{T}}^{*}\bm{H}\bm{F}_{n_{\mathrm{t}}}\bm{X},\label{spa}
\end{equation}
where $\bm{Y}_{n_{\mathrm{t}}}$ has dimension of $M_{\mathrm{T}}M_{\mathrm{RF}}\times N_{\mathrm{RF}}$, and $\bm{X}=\bm{I}_{N_{\mathrm{RF}}}$. Different from the AoD estimation, the receiver has full knowledge of the position of a specific receive auxiliary beam pair in $\bm{W}_{\mathrm{T}}$, e.g., $\left[\bm{W}_{\mathrm{T}}\right]_{:,p}=\bm{a}_{\mathrm{r}}(\eta_{m}-\delta_{\mathrm{r}})$, and $\left[\bm{W}_{\mathrm{T}}\right]_{:,q}=\bm{a}_{\mathrm{r}}(\eta_{m}+\delta_{\mathrm{r}})$, $p,q\in\left\{1,\cdots,M_{\mathrm{T}}M_{\mathrm{RF}}\right\}$. The corresponding ratio metric can therefore be computed as
\begin{eqnarray}\label{l12}
\zeta_{m,\ell}^{\mathrm{AoA}}=\frac{\left|\left[\bm{Y}_{n_{\mathrm{t}}}\right]_{p,\ell}\right|^{2}-\left|\left[\bm{Y}_{n_{\mathrm{t}}}\right]_{q,\ell}\right|^{2}}{\left|\left[\bm{Y}_{n_{\mathrm{t}}}\right]_{p,\ell}\right|^{2}+\left|\left[\bm{Y}_{n_{\mathrm{t}}}\right]_{q,\ell}\right|^{2}}=\frac{\left|\bm{a}_{\mathrm{r}}^{*}(\psi_{\ell})\bm{a}_{\mathrm{r}}(\eta_{m}-\delta_{\mathrm{r}})\right|^{2}-\left|\bm{a}_{\mathrm{r}}^{*}(\psi_{\ell})\bm{a}_{\mathrm{r}}(\eta_{m}+\delta_{\mathrm{r}})\right|^{2}}{\left|\bm{a}_{\mathrm{r}}^{*}(\psi_{\ell})\bm{a}_{\mathrm{r}}(\eta_{m}-\delta_{\mathrm{r}})\right|^{2}+\left|\bm{a}_{\mathrm{r}}^{*}(\psi_{\ell})\bm{a}_{\mathrm{r}}(\eta_{m}+\delta_{\mathrm{r}})\right|^{2}}
\end{eqnarray}
assuming $N_{\mathrm{tot}}M_{\mathrm{tot}}\rightarrow\infty$. The associated receive spatial frequency is estimated as
\begin{equation}\label{aoaangleR}
\hat{\psi}_{m,\ell}=\eta_{m}-\arcsin\left(\frac{\zeta_{m,\ell}^{\mathrm{AoA}}\sin(\delta_{\mathrm{t}})-\zeta_{m,\ell}^{\mathrm{AoA}}\sqrt{1-\left(\zeta_{m,\ell}^{\mathrm{AoA}}\right)^{2}}\sin(\delta_{\mathrm{r}})\cos(\delta_{\mathrm{r}})}{{\sin^{2}(\delta_{\mathrm{r}})+\left(\zeta_{m,\ell}^{\mathrm{AoA}}\right)^{2}\cos^{2}(\delta_{\mathrm{r}})}}\right).
\end{equation}
The corresponding receive array response vector can be constructed as $\bm{a}_{\mathrm{r}}(\hat{\phi}_{\ell})$ with $\hat{\phi}_{\ell}=\arcsin\big(\\ \lambda\hat{\psi}_{m,\ell}/2\pi d_{\mathrm{r}}\big)$. Finally, the receive array response matrix is constructed as $\hat{\bm{A}}_{\mathrm{r}}=\big[\bm{a}_{\mathrm{r}}(\hat{\phi}_{1}),\cdots,\\ \bm{a}_{\mathrm{r}}(\hat{\phi}_{N_{\mathrm{p}}})\big]$ accounting for all estimated AoAs.

In the absence of noise, the estimation performance of the proposed algorithm is only subject to the multi-path interference (see e.g., the second term in (\ref{rn})). By leveraging the high-power regime and channel sparsity, the multi-path interference can be minimized, and the corresponding ratio metric in (\ref{intratio}) (or (\ref{l12})) is independent of the analog receive (or transmit) probing. If the noise impairment is accounted for, (\ref{sp}) and (\ref{spa}) become to
\begin{eqnarray}
\bm{Y}_{m_{\mathrm{t}}}&=&\bm{W}_{m_{\mathrm{t}}}^{*}\bm{H}\bm{F}_{\mathrm{T}}\bm{X}+\bm{W}_{m_{\mathrm{t}}}^{*}\bm{P}, \label{nimt} \\ \bm{Y}_{n_{\mathrm{t}}}&=&\bm{W}_{\mathrm{T}}^{*}\bm{H}\bm{F}_{n_{\mathrm{t}}}\bm{X}+\bm{W}_{\mathrm{T}}^{*}\bm{Q}, \label{nint}
\end{eqnarray}
where $\bm{P}$ ($\bm{Q}$) is an $M_{\mathrm{RF}}\times N_{\mathrm{T}}N_{\mathrm{RF}}$ ($N_{\mathrm{T}}M_{\mathrm{RF}}\times N_{\mathrm{RF}}$) noise matrix given by concatenating $N_{\mathrm{T}}N_{\mathrm{RF}}$ ($N_{\mathrm{RF}}$) noise vectors. To estimate multi-path's AoD and AoA using (\ref{nimt}) and (\ref{nint}), the receive and transmit probing matrices $\bm{W}_{m'_{\mathrm{t}}}$ and $\bm{F}_{n'_{\mathrm{t}}}$ that satisfy $m'_{\mathrm{t}}=\underset{m_{\mathrm{t}}=1,\cdots,M_{\mathrm{T}}} {\mathrm{argmax}} ~\mathrm{tr}\left(\bm{Y}_{m_{\mathrm{t}}}^{*}\bm{Y}_{m_{\mathrm{t}}}\right)$ and $n'_{\mathrm{t}}=\underset{n_{\mathrm{t}}=1,\cdots,N_{\mathrm{T}}} {\mathrm{argmax}} ~\mathrm{tr}\left(\bm{Y}_{n_{\mathrm{t}}}^{*}\bm{Y}_{n_{\mathrm{t}}}\right)$ are first selected. By plugging $\bm{W}_{m'_{\mathrm{t}}}$ and $\bm{F}_{n'_{\mathrm{t}}}$ into (\ref{nimt}) and (\ref{nint}), the resulted $\bm{Y}_{m'_{\mathrm{t}}}$ and $\bm{Y}_{n'_{\mathrm{t}}}$ are then employed in (\ref{intratio}) and (\ref{l12}) to determine the ratio metrics. To efficiently execute this selection process, the steering angles of simultaneously probed beams should match the distribution of AoD/AoA, which are unknown to the transmitter and receiver in prior, as much as possible. Therefore, with finite $N_{\mathrm{T}}$ and $M_{\mathrm{T}}$, the analog beams in one probing matrix are steered to random angular directions in this paper. Using multiple transmit and receive RF chains, the number of attempts of the proposed algorithm then becomes to $N_{\mathrm{T}}N_{\mathrm{RF}}\times M_{\mathrm{T}}M_{\mathrm{RF}}$.

The pseudo-code of the proposed auxiliary beam pair enabled multi-path AoD estimation is provided in Algorithm 1.
\begin{algorithm}
  \caption{Multi-path AoD estimation via auxiliary beam pair design}
  \begin{algorithmic}
\State \textbf{Initialization}
\State 1:~Set the transmit and receive beam codebooks $\mathcal{F}_{\mathrm{T}}$ and $\mathcal{W}_{\mathrm{T}}$
\State 2:~Set the total numbers of transmit and receive probings $N_{\mathrm{T}}$ and $M_{\mathrm{T}}$
\State 3:~Set $\bm{F}_{n_{\mathrm{t}}}\in\mathbb{C}^{N_{\mathrm{tot}}\times N_{\mathrm{RF}}}$ ($n_{\mathrm{t}}=1,\cdots,N_{\mathrm{T}}$) such that $\left[\bm{F}_{n_{\mathrm{t}}}\right]_{:,i}$ ($i=1,\cdots,N_{\mathrm{RF}}$) is randomly\\
~~~chosen from $\mathcal{F}_{\mathrm{T}}$
\State 4:~Set $\bm{W}_{m_{\mathrm{t}}}\in\mathbb{C}^{M_{\mathrm{tot}}\times M_{\mathrm{RF}}}$ ($m_{\mathrm{t}}=1,\cdots,M_{\mathrm{T}}$) such that $\left[\bm{W}_{m_{\mathrm{t}}}\right]_{:,\kappa}$ ($\kappa=1,\cdots,M_{\mathrm{RF}}$) is randomly\\
~~~chosen from $\mathcal{W}_{\mathrm{T}}$
\State 5:~Set $\bm{F}_{\mathrm{T}}=\left[\bm{F}_{1},\cdots,\bm{F}_{n_{\mathrm{t}}},\cdots,\bm{F}_{N_{\mathrm{T}}}\right]$ and $\bm{W}_{\mathrm{T}}=\left[\bm{W}_{1},\cdots,\bm{W}_{m_{\mathrm{t}}},\cdots,\bm{W}_{M_{\mathrm{T}}}\right]$
\State \textbf{Find the best receive combining matrix}
\State 6:~~\textbf{For}~$m_{\mathrm{t}}=1,\cdots,M_{\mathrm{T}}$
\State 7:~~~~\textbf{For}~$n_{\mathrm{t}}=1,\cdots,N_{\mathrm{T}}$
\State 8:~~~~~~$\left[\bm{Y}_{m_{\mathrm{t}}}\right]_{:,N_{\mathrm{RF}}\left(n_{\mathrm{t}}-1\right)+1:N_{\mathrm{RF}}n_{\mathrm{t}}}=\bm{W}_{m_{\mathrm{t}}}^{*}\bm{H}\bm{F}_{n_{\mathrm{t}}}\bm{X}+\bm{W}_{m_{\mathrm{t}}}^{*}\left[\bm{P}\right]_{:,N_{\mathrm{RF}}\left(n_{\mathrm{t}}-1\right)+1:N_{\mathrm{RF}}n_{\mathrm{t}}}$
\State 9:~~~~~\textbf{end~For}
\State 10:~~\textbf{end~For}
\State 11:~~$m'_{\mathrm{t}}=\underset{m_{\mathrm{t}}=1,\cdots,M_{\mathrm{T}}} {\mathrm{argmax}} ~\mathrm{tr}\left(\bm{Y}_{m_{\mathrm{t}}}^{*}\bm{Y}_{m_{\mathrm{t}}}\right)$ $\Rightarrow$ $\bm{W}_{m'_{\mathrm{t}}}$ is selected for the rest of the procedures
\State \textbf{Find the best transmit auxiliary beam pair for path-$\ell$, $\ell=1,\cdots,N_{\mathrm{p}}$}
\State 12:~~\textbf{For}~$n'=1,\cdots,N_{\mathrm{K}}$
\State 13:~~~~Determine $u'$ and $v'$ ($u',v'\in\left\{1,\cdots,N_{\mathrm{T}}N_{\mathrm{RF}}\right\}$) such that $\left[\bm{F}_{\mathrm{T}}\right]_{:,u'}=\bm{a}_{\mathrm{t}}(\nu_{n'}-\delta_{\mathrm{t}})$, and \\ ~~~~~~~$\left[\bm{F}_{\mathrm{T}}\right]_{:,v'}=\bm{a}_{\mathrm{t}}(\nu_{n'}+\delta_{\mathrm{t}})$ by detecting the beam ID
\State 14:~~~~Calculate $\left|\left[\bm{Y}_{m'_{\mathrm{t}}}\right]_{\ell,u'}\right|^2$ and $\left|\left[\bm{Y}_{m'_{\mathrm{t}}}\right]_{\ell,v'}\right|^2$
\State 15:~~\textbf{end~For}
\State 16:~~$\left[\bm{F}_{\mathrm{T}}\right]_{:,u}$ and $\left[\bm{F}_{\mathrm{T}}\right]_{:,v}$ are chosen as the auxiliary beam pair of interest using Lemma 2
\State 17:~Determine $n\in\left\{1,\cdots,N_{\mathrm{K}}\right\}$ such that $\left[\bm{F}_{\mathrm{T}}\right]_{:,u}=\bm{a}_{\mathrm{t}}(\nu_{n}-\delta_{\mathrm{t}})$, and $\left[\bm{F}_{\mathrm{T}}\right]_{:,v}=\bm{a}_{\mathrm{t}}(\nu_{n}+\delta_{\mathrm{t}})$
\State \textbf{Estimate path-$\ell$'s AoD}
\State 18:~Compute $\zeta_{n,\ell}^{\mathrm{AoD}}$ following (\ref{intratio}) and $\hat{\mu}_{n,\ell}$ using (\ref{aodangleR})
  \end{algorithmic}
\end{algorithm}

\section{Applications of Auxiliary Beam Pair Design in MmWave Cellular Systems}
Potential deployment scenarios of the proposed design approach in mmWave cellular systems are addressed in this section, including control channel beamforming and extension to the multi-user scenario.
\subsection{Analog-only beamforming for control channel} To achieve sufficiently high received signal quality, we believe that the control channel will be beamformed in mmWave cellular systems. We classify the control channel as system-specific, cell-specific and user-specific control channel due to their different purposes and link-budget requirements. For instance, the system-specific control channel carries basic network information such as operating carrier frequency, bandwidth, and etc., which needs a wide coverage. The cell-specific control channel conveys the signals that help the user equipment (UE) to discover, synchronize and access the cell, which requires not only a certain level of coverage, but also relatively high received signal strength. For user-specific control channel, which embeds reference signals for channel information acquisition for a specific UE, \emph{pencil} beams are employed to ensure high received signal quality. A multi-layer structure is proposed in this paper for control channel beamforming such that analog transmit beams in different layers are associated with different types of control channel. In Fig.~6, examples of the multi-layer control channel beamforming are presented. Layer-$1$, $2$ and $3$ transmit system-specific, cell-specific and user-specific control channels using coarse and fine-grained beams.
\begin{figure}
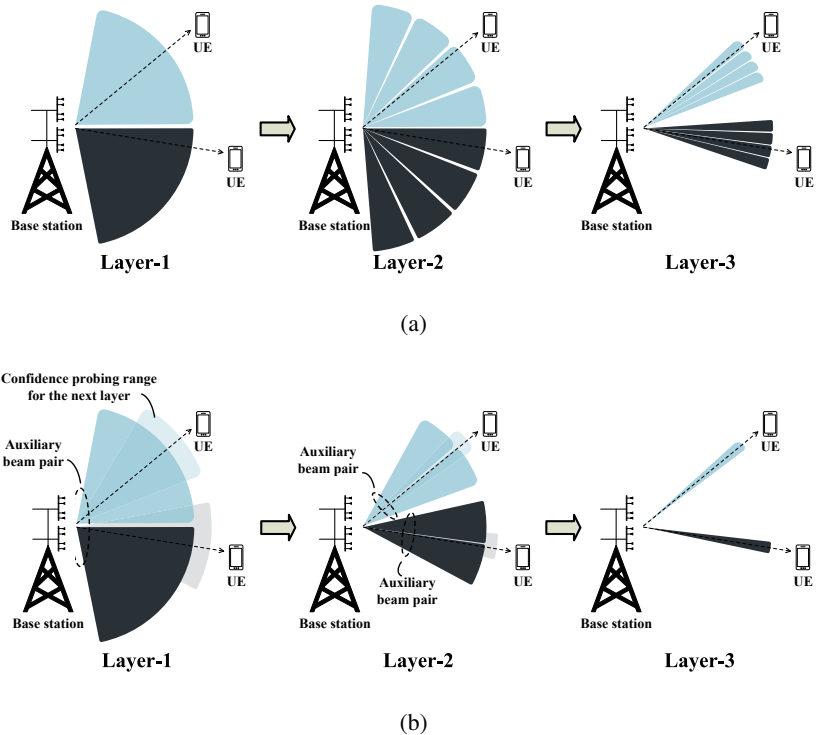

\centering
\subfigure[]{%
\includegraphics[width=4.365in]{control_channel.pdf}
\label{fig:subfigure1}}
\quad
\subfigure[]{%
\includegraphics[width=4.35in]{control_channel_ABP.pdf}
\label{fig:subfigure2}}
\caption{(a) Multi-layer grid-of-beams based control channel beamforming in mmWave cellular systems. (b) Multi-layer auxiliary beam pair based control channel beamforming in mmWave cellular systems.}
\label{fig:figure}
\end{figure}

One example of a grid-of-beams based multi-layer control channel beamforming design is provided in Fig.~6(a). As the grid-of-beams based approach is not able to provide high-resolution estimates of channel's AoD and AoA, each UE in a given layer searches over all analog transmit beams within the probing range of interest, and selects the one with the highest received signal strength. Using the example shown in Fig.~6(a), the total number of attempts for a given UE to finally select the user-specific control channel beam is $2+4+4=10$.

Fig.~6(b) exhibits one example of employing auxiliary beam pairs in multi-layer beamformed control channel design. In each layer, the analog transmit beams not only carry necessary control signals, but also act as auxiliary beam pair to help acquire channel information. For instance, in Layer-$1$, one auxiliary beam pair is formed, in which two $60^{\circ}$ analog transmit beams are contained covering a $120^{\circ}$ sector. Upon receiving the analog transmit beams, each UE not only decodes the system-specific information from the beam that yields the highest received signal strength, but also calculates the ratio metric corresponding to the auxiliary beam pair. The ratio metric is then quantized and sent back to the base station (BS). According to the ratio metric feedback, the BS estimates the AoD and determines the confidence probing range for the next layer beamforming. For instance, assume that $\hat{\mu}_{\ell_{1}}$ is the estimated transmit spatial frequency obtained via Layer-$1$ beamforming, and $\delta_{\ell_{2}}$ is half of the beamforming range of the analog transmit beam used for Layer-$2$ beamforming. The confidence probing range for Layer-$2$ is then determined as $\left[\hat{\mu}_{\ell_{1}}-\delta_{\ell_{2}},\hat{\mu}_{\ell_{1}}+\delta_{\ell_{2}}\right]$. The above process repeats until the final layer beamforming has been executed. Using the example shown in Fig.~6(b), the total number of attempts for a given UE to finally select the user-specific control channel beam becomes to $2+2+2=6$. In contrast to the grid-of-beams based approach, the beam finding overhead is reduced by $40\%$.
\subsection{Extension to multi-user scenario}
In previous sections, the proposed AoD/AoA estimation algorithm is specifically illustrated for a single-user scenario. Extension of the proposed approach to a multi-user setup is discussed in this part as this application scenario is important for practical cellular systems.

Assume perfect synchronization between the BS and UEs, the algorithms proposed for the single-user case can be directly extended to the multi-user scenario with appropriate frame structure design that can better support the communications between the BS and UEs through auxiliary beam pairs. A probing frame is therefore defined which includes a predetermined number of probing slots. From the perspective of each UE, the same receive probing is performed for all probing slots. The receive probing can be either single receive beam if the UE is equipped with single RF chain, or receive combining matrix if multiple RF chains are employed. For a given probing frame, all transmit probings are conducted by the BS across all probing slots in the probing frame. That is, in a given probing frame, the UE uses the same receive probing to combine all transmit probings across all probing slots, and this procedure continues until all receive probings are executed by the UE across all probing frames. Following steps 6$\sim$18 in Algorithm 1, each UE then estimates the desired AoDs/AoAs. Note that for the multi-user scenario, the auxiliary beam pair for angle estimation plays a similar role to the common reference signal (CRS) in LTE for channel estimation such that the probed beam-specific signals are common for all active users.

\section{Numerical Results}
In this section, we evaluate the performance of the proposed auxiliary beam pair enabled channel estimation technique. Both the transmitter and receiver employ a ULA with inter-element spacing $\lambda/2$, and $180^{\circ}$ boresight. The codebook for quantizing the ratio measure in the proposed method has non-uniformly distributed codewords within the interval of $[-1,1]$ as derived in Section III. Denote by $N_{\mathrm{ant}}=N_{\mathrm{tot}}$ (or $M_{\mathrm{tot}}$), and $\delta=\delta_{\mathrm{t}}$ (or $\delta_{\mathrm{r}}$) for AoD (or AoA) estimation. In the simulation, we set $\delta=(\pi/2)/N_{\mathrm{ant}}$. For $\left\{N_{\mathrm{ant}}=4,\delta=\pi/8\right\}$, $\left\{N_{\mathrm{ant}}=8,\delta=\pi/16\right\}$ and $\left\{N_{\mathrm{ant}}=16,\delta=\pi/32\right\}$, the corresponding codebook size of transmit/receive beams is $4$, $10$, and $20$. Denote half of the $3$dB beamwidth for a ULA with half-wavelength antenna spacing by $\vartheta_{3\mathrm{dB}}$. We therefore have $\vartheta_{3\mathrm{dB}}$ approximated as $102^{\circ}/N_{\mathrm{ant}}$ \cite{antennabook}. We set $\delta\lessapprox\vartheta_{3\mathrm{dB}}$ to ensure seamless coverage. For instance, for $N_{\mathrm{ant}}=4$, the codebook size is $180^{\circ}/2\delta=180^{\circ}/45^{\circ}=4$.
\begin{figure}
\centering
\subfigure[]{%
\includegraphics[width=2.95in]{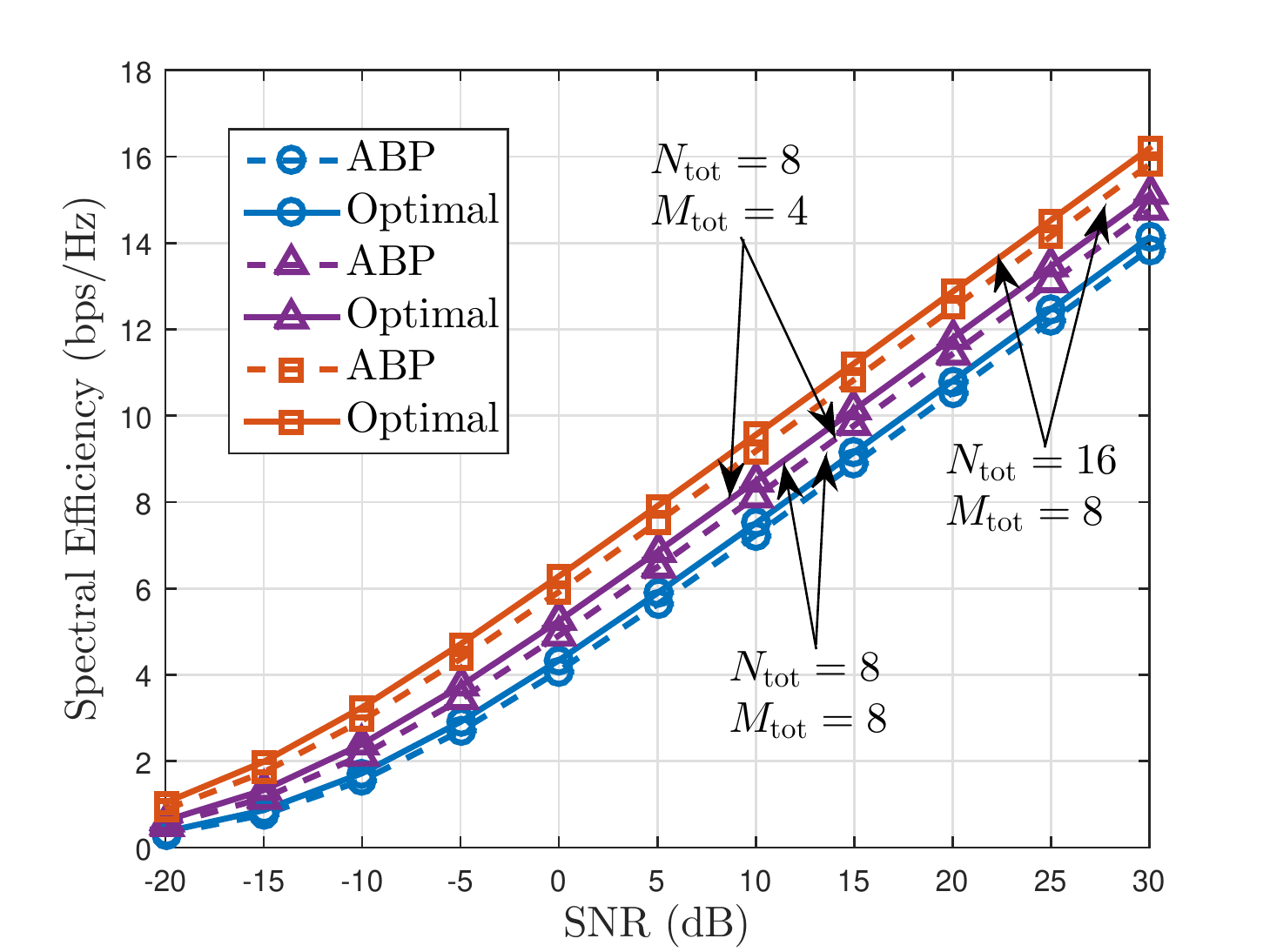}
\label{fig:subfigure3}}
\hspace{-10mm}
\subfigure[]{%
\includegraphics[width=2.95in]{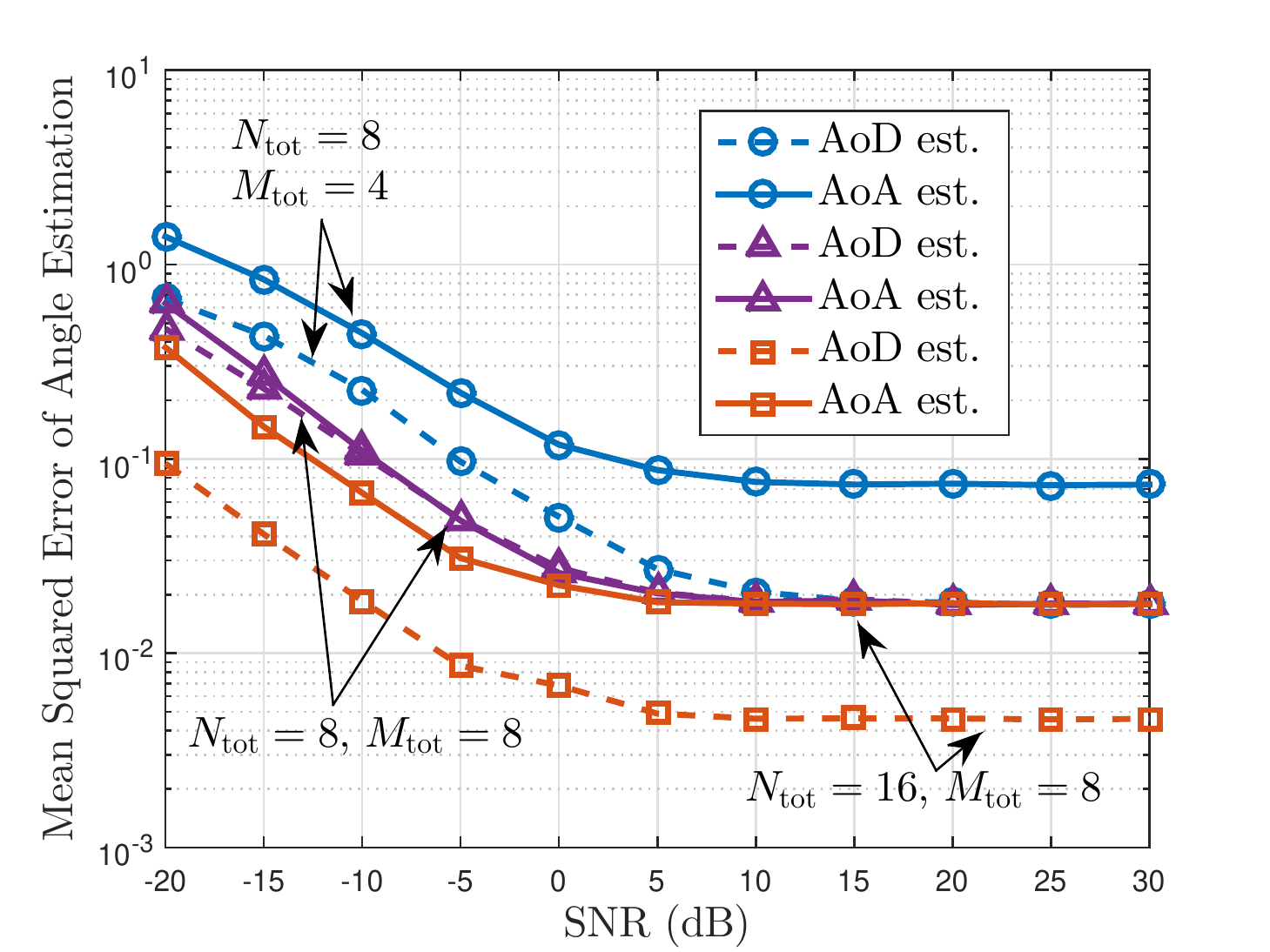}
\label{fig:subfigure3}}
\caption{(a) Spectral efficiency of analog-only beamforming and combining using perfect channel information and estimated AoD and AoA via auxiliary beam pair design in a single-path channel. (b) Mean squared error performance of AoD and AoA estimation using the proposed approach under various SNR levels.}
\label{fig:figure}
\end{figure}

In Fig.~7(a), the spectral efficiency performance of analog-only beamforming and combining is evaluated under single-path channel conditions. Both the AoD and AoA are uniformly distributed in the interval of $\left[-\pi/2,\pi/2\right]$ for each channel use. The transmit beamforming and receive combining vectors are constructed using (i) perfect channel knowledge of AoD and AoA, and (ii) estimated AoD and AoA via auxiliary beam pair design. The ratio measure that characterizes the AoD for each channel use is quantized using $4$ bits. It can be observed from Fig.~7(a) that the spectral efficiency performance of employing the proposed channel estimation method approaches that with perfect channel knowledge, for different numbers of employed transmit antennas. The MSE performance of the single-path's AoD and AoA estimation is provided in Fig.~7(b) under various SNR levels. It is observed from Fig.~7(b) that promising MSE performance of AoD and AoA estimation can be achieved even in a relatively low SNR regime. With an increase in the number of employed antennas, the AoD and AoA estimation performance is further improved.

\begin{figure}
\centering
\subfigure[]{%
\includegraphics[width=2.95in]{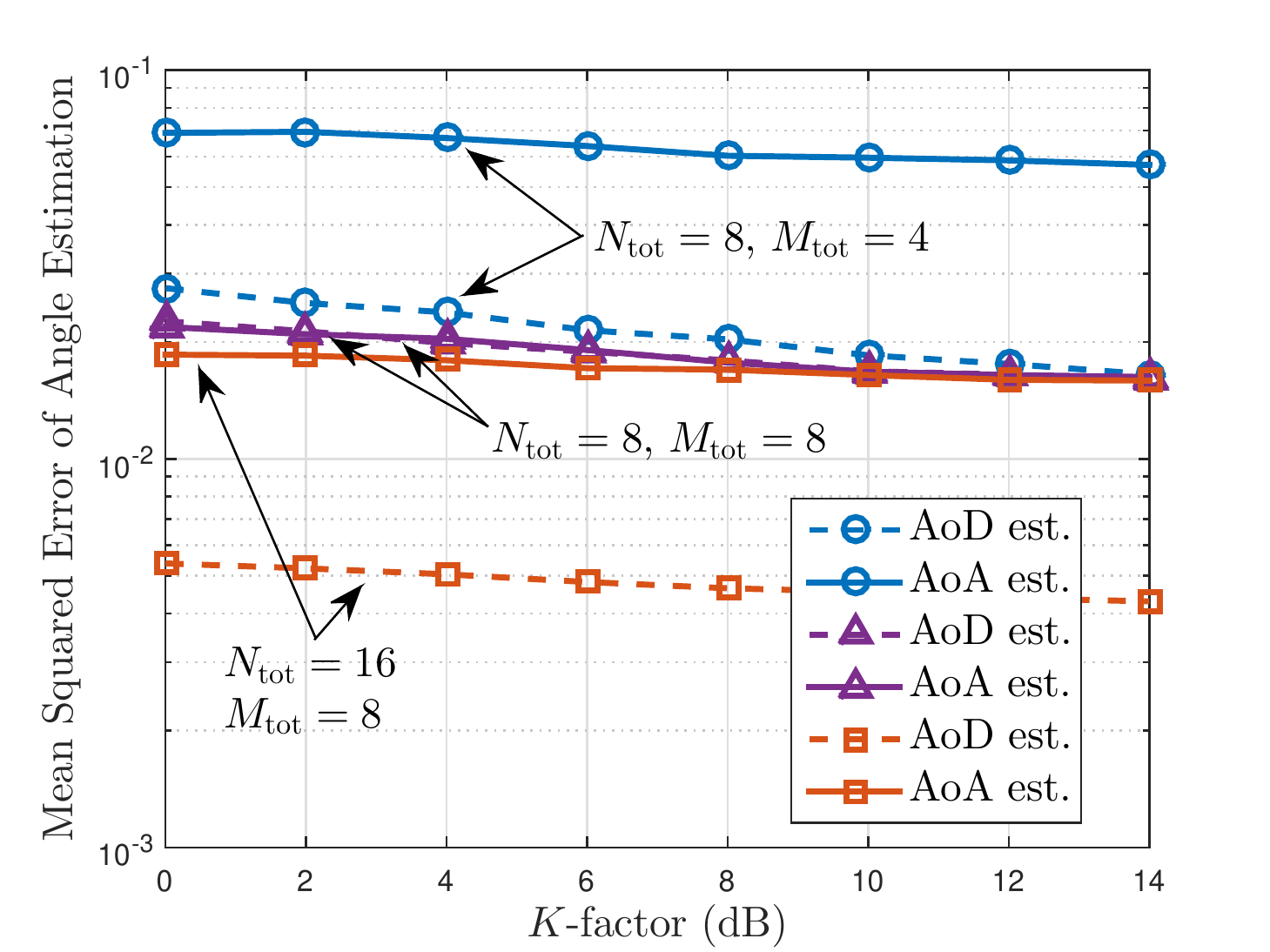}
\label{fig:subfigure3}}
\hspace{-10mm}
\subfigure[]{%
\includegraphics[width=2.95in]{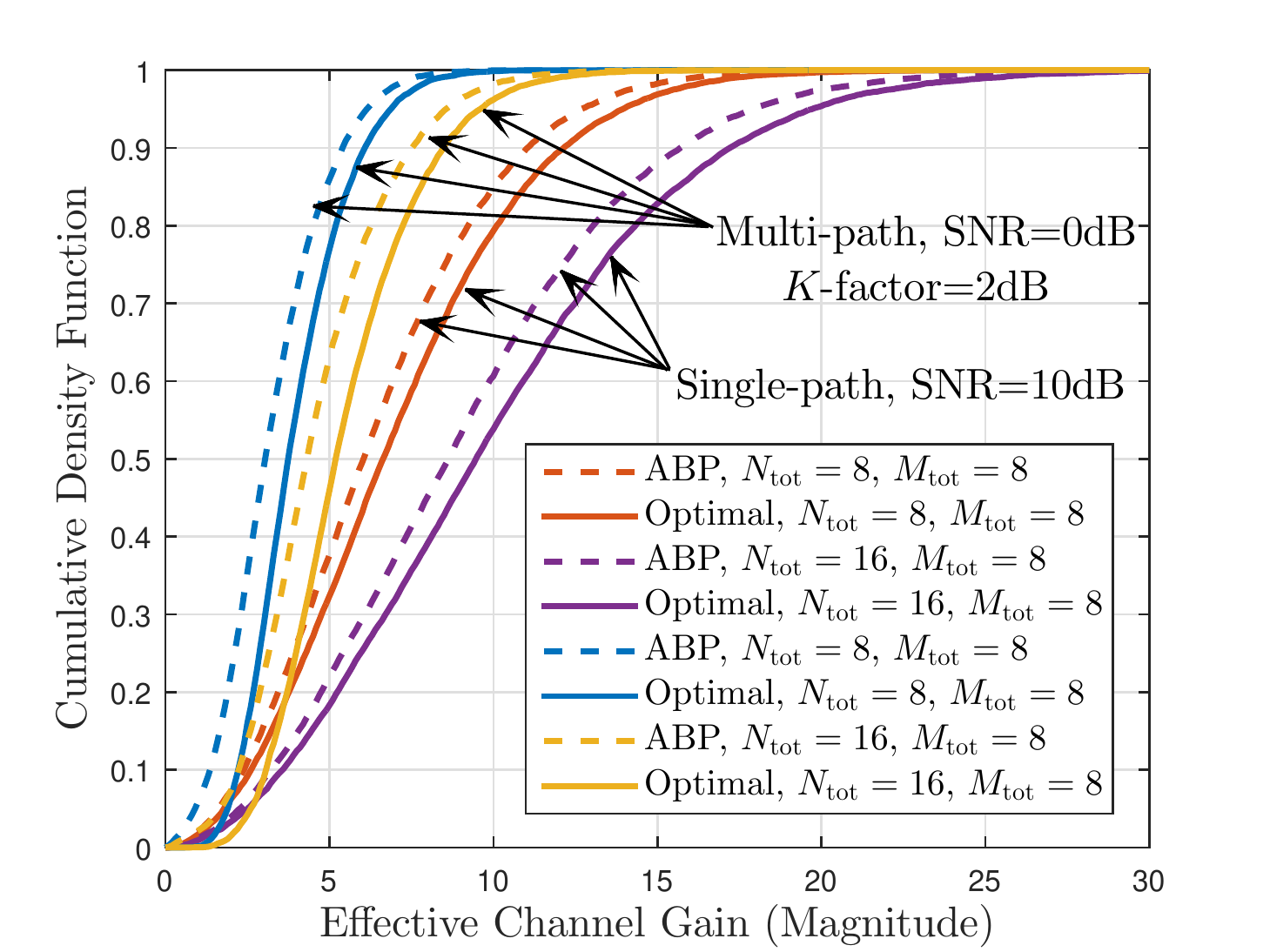}
\label{fig:subfigure3}}
\caption{(a) Mean squared error performance of AoD and AoA estimation using the proposed approach under various $\mathsf{K}$-factor values. (b) Cumulative density function of effective channel gain of analog-only beamforming and combing using perfect channel information and estimated AoD and AoA via auxiliary beam pair design.}
\label{fig:figure}
\end{figure}
The corresponding MSE performance of AoD and AoA estimation is plotted in Fig.~8(a) under multi-path channel condition and $10$dB SNR. In this example, a Rician channel model is assumed, with various Rician $\mathsf{K}$-factor values,
\begin{align}
\bm{H} = \sqrt{\frac{\mathsf{K}}{1+\mathsf{K}}}\underbrace{\alpha_{\ell}\bm{a}_{\mathrm{r}}(\phi_{\ell})\bm{a}_{\mathrm{t}}^{*}(\theta_{\ell})}_{\bm{H}_{\mathrm{LOS}}}+\sqrt{\frac{\mathsf{1}}{1+\mathsf{K}}}\underbrace{\sum_{\ell'=1,r'\neq r}^{N_{\mathrm{p}}}\alpha_{\ell'}\bm{a}_{\mathrm{r}}(\phi_{\ell'})\bm{a}_{\mathrm{t}}^{*}(\theta_{\ell'})}_{\bm{H}_{\mathrm{NLOS}}},
\end{align}
where $\bm{H}_{\mathrm{LOS}}$ and $\bm{H}_{\mathrm{NLOS}}$ represent line-of-sight (LOS) and non-LOS (NLOS) channel components. The number of NLOS channel components is $5$. The objective is to estimate the dominant path's AoD and AoA, and steer the analog transmit and receive beams towards the estimated angles. From the channel observation in \cite{rician}, $13.2$dB Rician $\mathsf{K}$-factor characterizes the mmWave channels' conditions the best in an urban wireless channel topography. The reason for evaluating various Rician $\mathsf{K}$-factor values is to validate the capability of the proposed algorithm in estimating the dominant path's AoD/AoA when NLOS components have relatively strong power. With directional analog beamforming and combining, the MSE performance of AoD and AoA estimation is still promising even in the small $\mathsf{K}$-factor regime (e.g., $2$dB). In Fig.~8(b), the effective channel gain $\left|\bm{w}^{*}\bm{H}\bm{f}\right|^2$ is evaluated, where $\bm{f}$ and $\bm{w}$ are analog transmit and receive steering vectors obtained using perfect channel information and estimated AoD and AoA via auxiliary beam pair design. It can be observed from Fig.~8(b) that the proposed technique approaches that with perfect channel knowledge.

\begin{figure}
\centering
\subfigure[]{%
\includegraphics[width=2.95in]{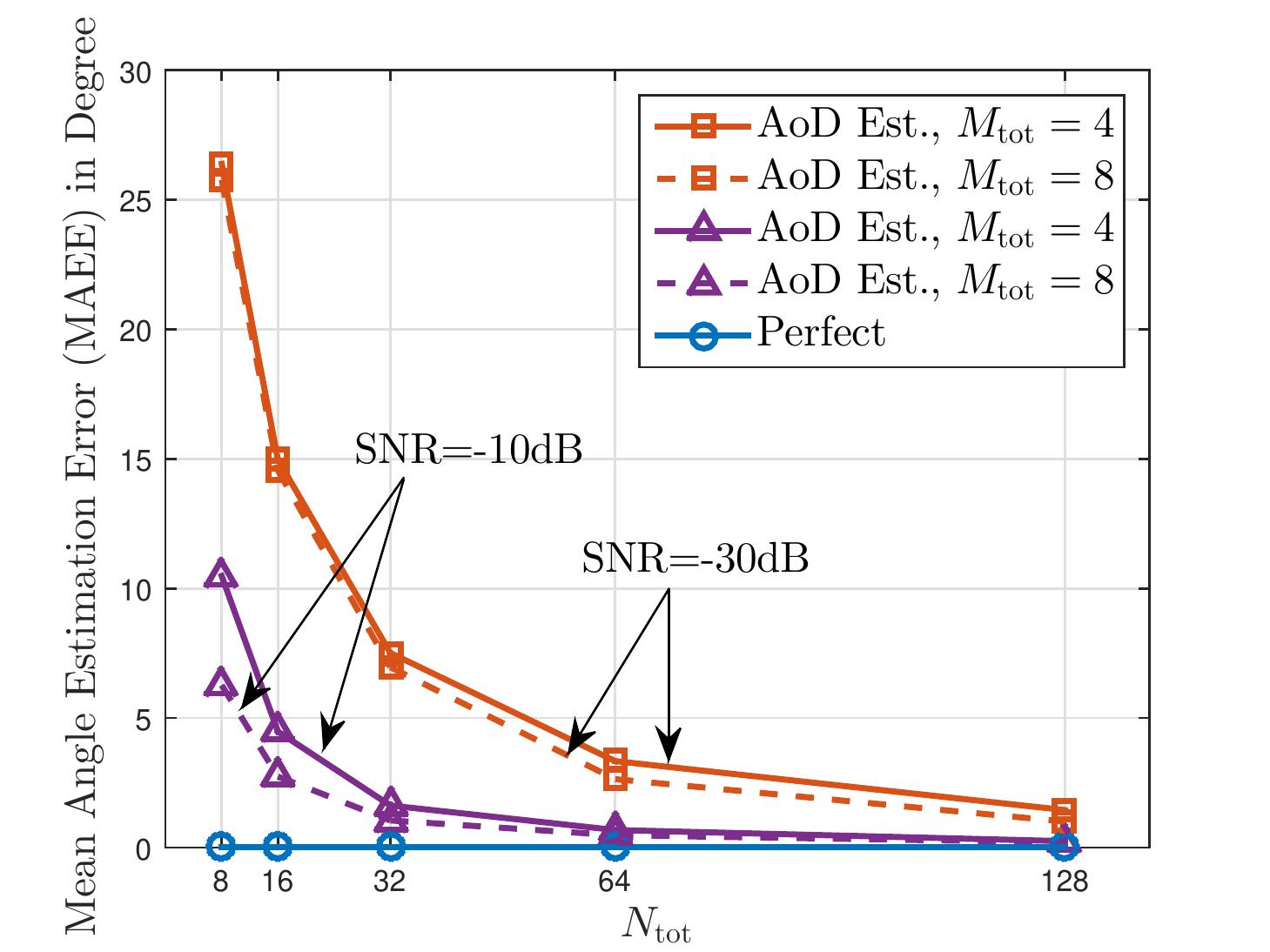}
\label{fig:subfigure3}}
\hspace{-8mm}
\subfigure[]{%
\includegraphics[width=2.95in]{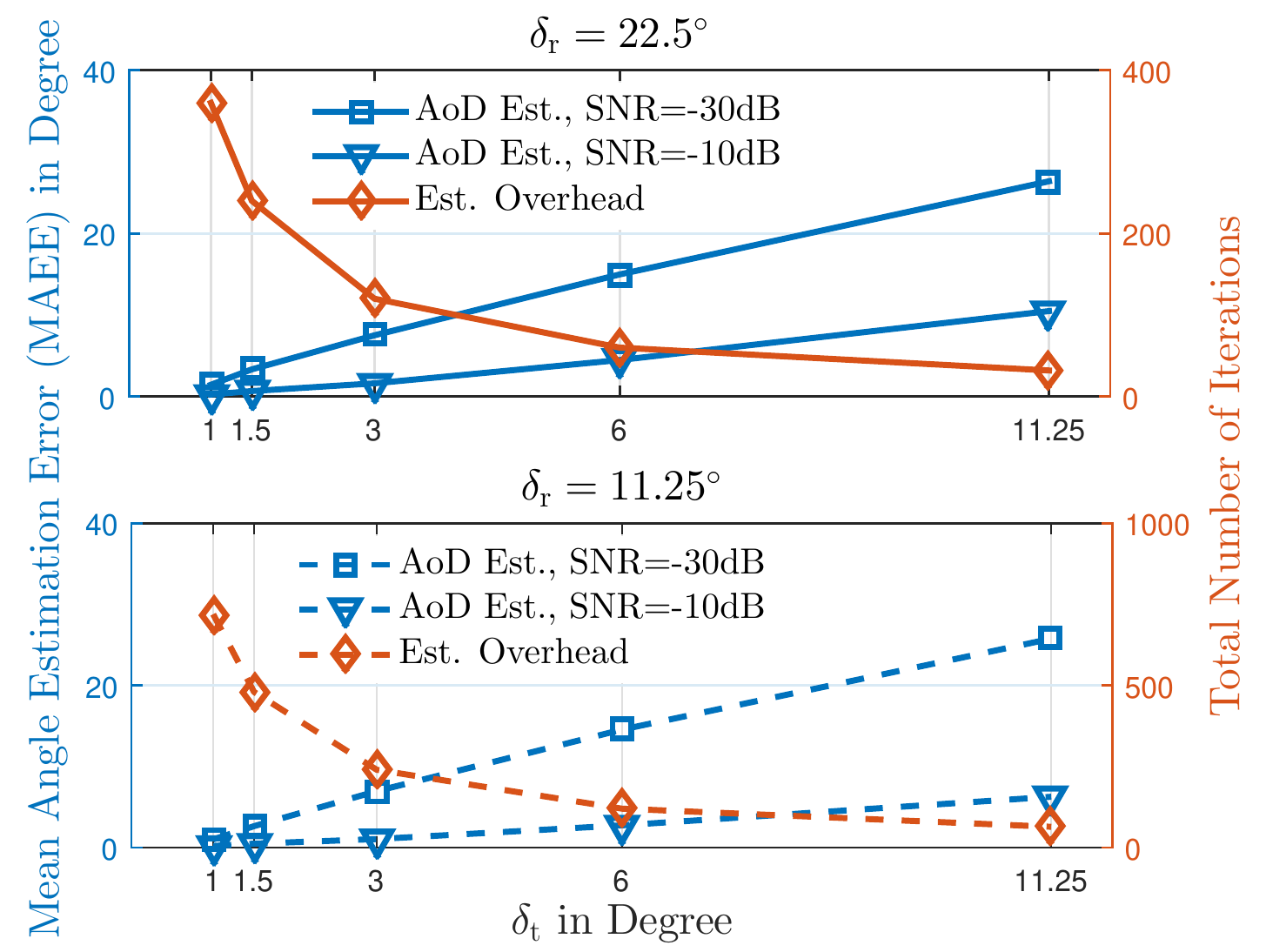}
\label{fig:subfigure3}}
\caption{(a) Mean angle estimation error (MAEE) of AoD estimation using the proposed approach with different numbers of transmit antennas. (b) Trade-off performance between the angle estimation and the total number of iterations between the transmitter and receiver.}
\label{fig:figure}
\end{figure}
In Figs.~7 and 8, the performance of the proposed design approach is evaluated assuming relatively small numbers of transmit and receive antennas, and a wide range of target SNRs. The mmWave systems, however, are expected to deploy a large number of antennas to provide sufficient link margin via directional beamforming. In Fig.~9(a), the mean angle estimation error (MAEE) performance is evaluated for the proposed algorithm using different numbers of transmit antennas in the single-path channel. Here, the MAEE is defined as $\mathbb{E}\left[\left|\upsilon_{\mathrm{true}}-\upsilon_{\mathrm{est}}\right|\right]$, where $\upsilon_{\mathrm{true}}$ represents the exact angle in degree, and $\upsilon_{\mathrm{est}}$ is its estimated counterpart in degree. From Fig.~9(a), it can be observed that with an increase in the number of transmit antennas, the MAEE is reduced for both $-10$dB and $-30$dB SNRs. For large $N_{\mathrm{tot}}M_{\mathrm{tot}}$, e.g., $128\times 8$ in this example, the MAEE is close to $0^\circ$ for different SNR levels. The estimation performance improvements, however, are brought by increased estimation overhead. In Fig.~9(b), the trade-off performance between the estimation overhead in terms of the total number of iterations between the transmitter and receiver, and the angle estimation performance determined by $\delta$ is evaluated. For instance, $\delta=11.25^{\circ}$ corresponds to $N_{\mathrm{ant}}=8$, and eight beams are formed to cover the $180^{\circ}$ angular range. That is, for $\delta_{\mathrm{t}}=\delta_{\mathrm{r}}=11.25^{\circ}$, the total number of iterations are calculated as $8\times 8=64$, and the MAEE performance is obtained by setting $N_{\mathrm{tot}}=M_{\mathrm{tot}}=8$ for a given target SNR (as shown in Fig.~9(a)). From the link-level evaluation results shown in Fig.~9(b), it can be concluded that due to the existence of the performance trade-off, appropriate $\delta$ can be determined for a given number of iterations between the transmitter and receiver.

\begin{figure}
\centering
\subfigure[]{%
\includegraphics[width=2.95in]{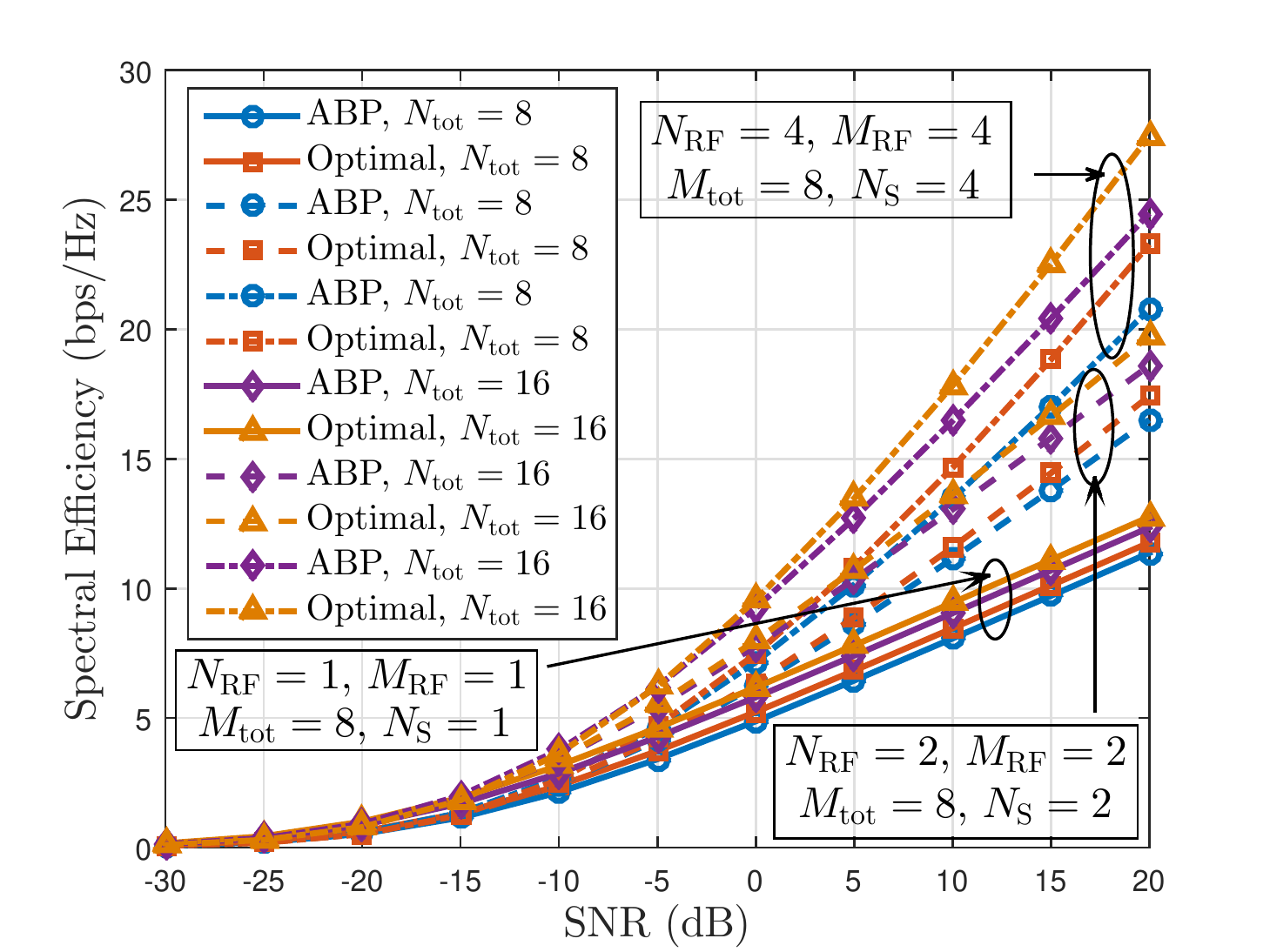}
\label{fig:subfigure3}}
\hspace{-10mm}
\subfigure[]{%
\includegraphics[width=2.95in]{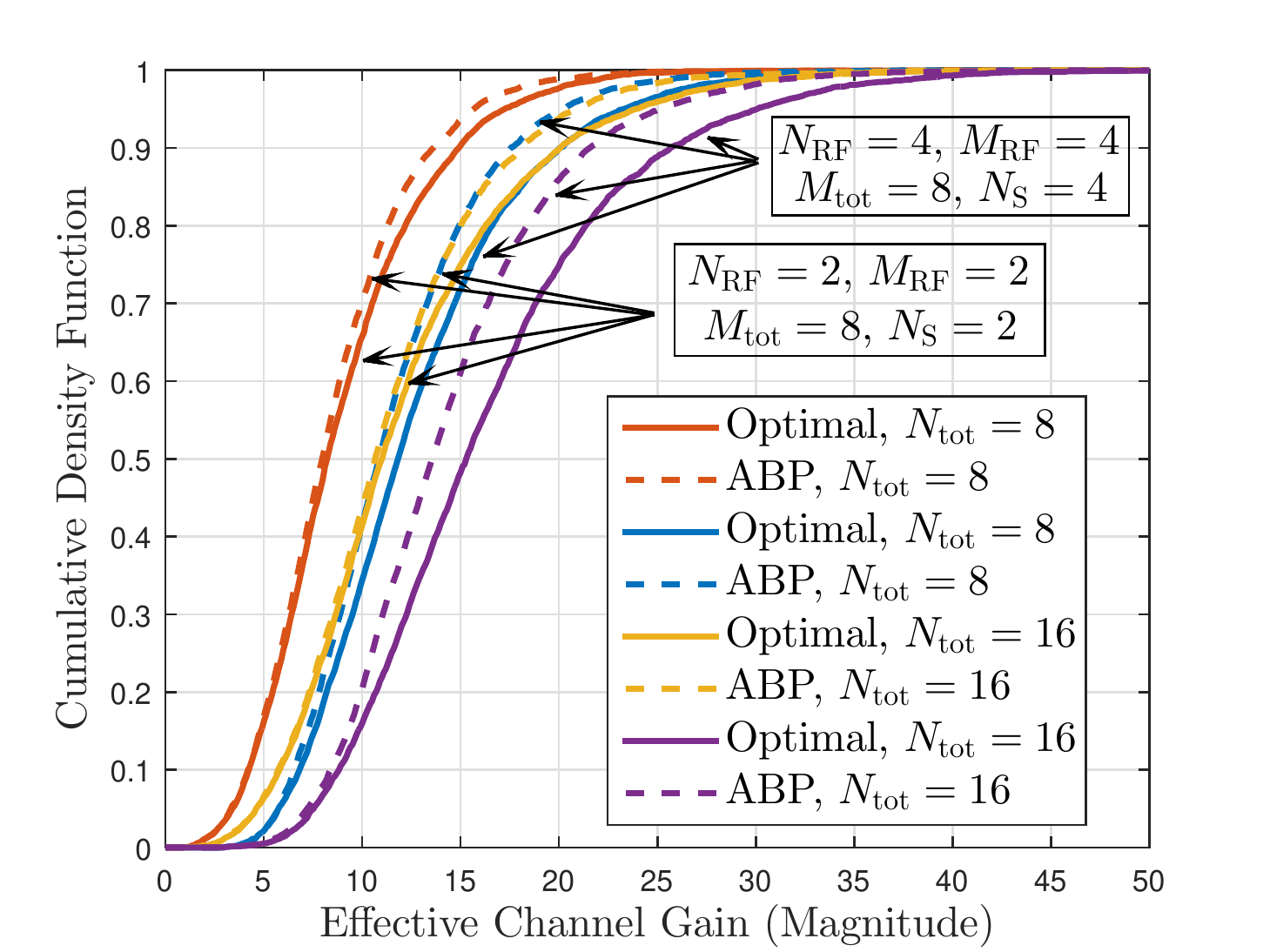}
\label{fig:subfigure3}}
\caption{(a) Spectral efficiency of hybrid analog and digital precoding using perfect channel information and estimated AoDs and AoAs via auxiliary beam pair design in a multi-path channel. (b) Cumulative density function of effective channel gain of hybrid precoding using perfect channel information and estimated AoDs and AoAs via auxiliary beam pair design.}
\label{fig:figure}
\end{figure}
In Fig.~10(a), multi-stream transmission via hybrid analog and digital precoding is depicted in terms of the spectral efficiency. The channel model described in (\ref{chm}) with $N_{\mathrm{p}}=\min(N_{\mathrm{RF}},M_{\mathrm{RF}})$ is employed, in which the AoD and AoA are assumed to take continuous values, i.e., not quantized, and are uniformly distributed in $\left[-\pi/2,\pi/2\right]$. The digital baseband precoder is selected from either $2\times 2$ or $4\times 4$ codebook employed in the LTE standard \cite{lte}. The total number of successive probings at the transmitter and receiver are set to be $N_{\mathrm{T}}=30$ and $M_{\mathrm{T}}=20$. It is observed from Fig.~10(a) that using the proposed approach to estimate multi-path components, the resulting spectral efficiency performance is almost identical to the optimal digital precoding and combining assuming perfect channel knowledge at relatively low SNR regime. The associated effective channel gain of hybrid MIMO is plotted in Fig.~10(b) assuming $0$dB SNR, from which similar observations are obtained to Fig.~8(b).

\begin{figure}
\begin{center}
\includegraphics[width=3.35in]{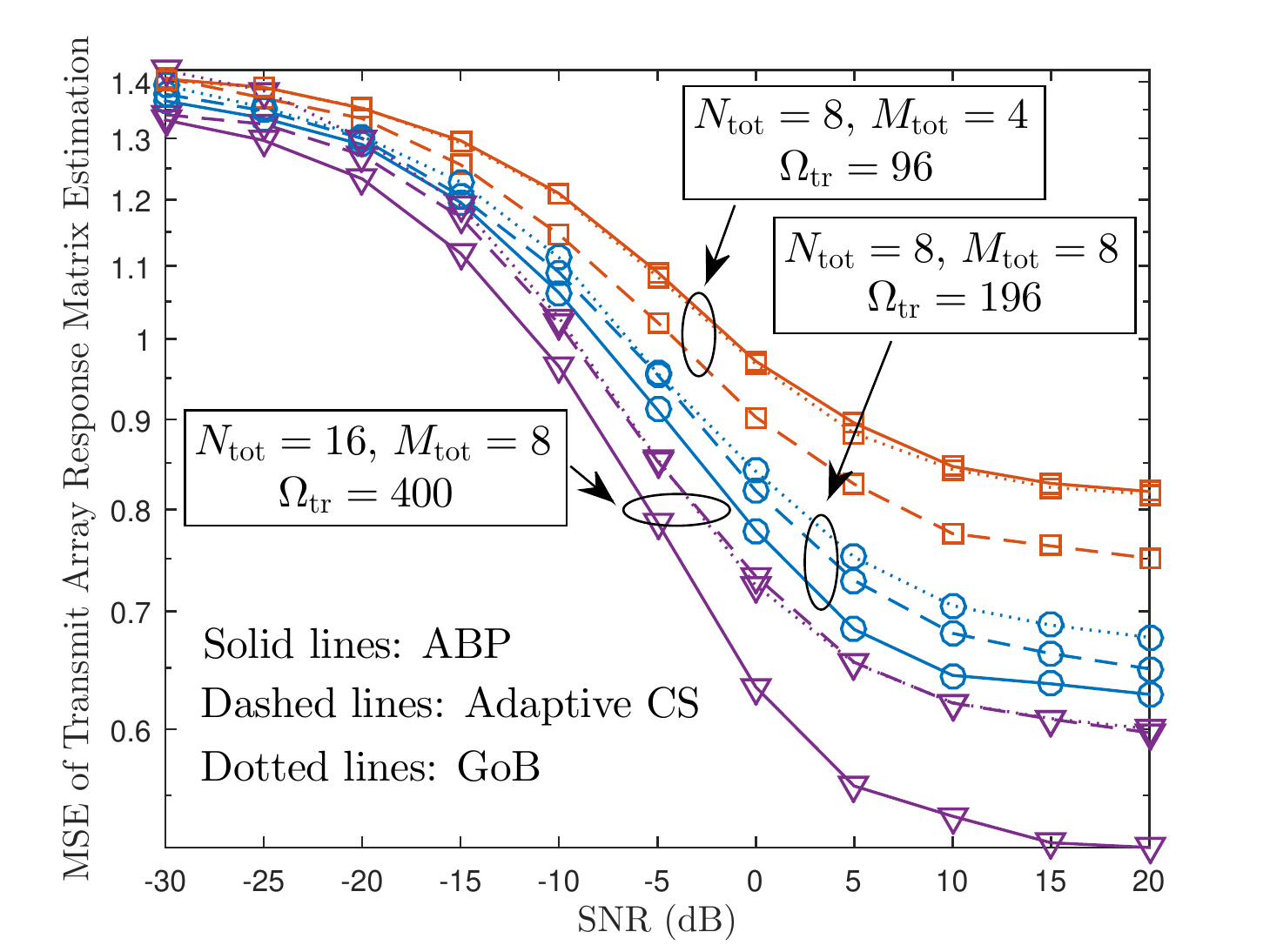}
\caption{MSE performance of transmit array response matrix estimation using the proposed approach, adaptive CS \cite{ahmedce} and modified GoB \cite{singh2} versus various SNR values.}
\end{center}
\end{figure}
In Fig.~11, the MSE performance of transmit array response matrix estimation is evaluated using the proposed design approach, adaptive compressed sensing (CS) technology developed in \cite{ahmedce} and modified grid-of-beams (GoB) based approach in \cite{singh2} using the proposed probing strategy in section III-E. The performance metric is defined as $\mathbb{E}\left[\mathrm{tr}\left(\bm{\Gamma}^{*}\bm{\Gamma}\right)\right]$, where $\bm{\Gamma}=\bm{A}_{\mathrm{t}}-\hat{\bm{A}}_{\mathrm{t}}$, and $\hat{\bm{A}}_{\mathrm{t}}$ contains the estimated transmit array response vectors. Here, $N_{\mathrm{p}}=N_{\mathrm{RF}}=M_{\mathrm{RF}}=3$. The total number of \emph{training steps} $\Omega_{\mathrm{tr}}$ defined in \cite{ahmedce} is interpreted as the total number of probings in the proposed approach and modified GoB. For fair comparison, we set $\Omega_{\mathrm{tr}}=96, 196$ and $400$. These correspond to $N_{\mathrm{T}}\times M_{\mathrm{T}}=12\times8, 14\times14$ and $20\times20$. From Fig.~11, it is observed that with $M_{\mathrm{tot}}=4$, the adaptive CS approach exhibits better MSE performance than the proposed approach. The reason is that with relatively small set of candidate receive combining vectors and $M_{\mathrm{T}}$, the proposed approach is not able to provide sufficient receive array gain to achieve high-resolution angle estimate. With an increase in the number of receive antennas and total number of probings, however, the proposed design approach outperforms the adaptive CS and GoB based methods. The reason is that the performance of the adaptive CS and GoB is limited by the codebook resolution, while the proposed approach is able to provide super-resolution AoD estimate as long as multi-path components can be resolved. It can also be observed from Fig.~11 that, the auxiliary beam pair based design with $N_{\mathrm{tot}}=8$ and $M_{\mathrm{tot}}=8$ exhibits similar angle estimation performance to the grid-of-beams based method with $N_{\mathrm{tot}}=16$ and $M_{\mathrm{tot}}=8$ for various SNR values. Fewer transmit antennas implies that fewer wider beams are required to cover a given angular range, which in turn, reduces the number of iterations between the transmitter and receiver.

\section{Conclusion}
In this paper, we proposed novel auxiliary beam pair design to perform channel parameters estimation in closed-loop mmWave systems with large number of deployed antennas. By leveraging well structured pairs of custom designed analog beams, high-resolution estimates of channel's AoD and AoA can be obtained via amplitude comparison. Using the estimated AoD and AoA, the directional initial access process can be facilitated, and data channel spatial multiplexing can be enabled via hybrid precoding. Numerical results reveal that by employing the proposed algorithm, promising channel estimation performance is achieved with significantly reduced training overhead under a moderate amount of feedback.
\appendix
\begin{center}
\emph{Proof of Lemma 3}
\end{center}
For a given analog transmit beam $\bm{a}_{\mathrm{t}}(\nu_{n}+\delta_{\mathrm{t}})$, the received signals after analog combining with $\bm{a}_{\mathrm{r}}(\eta_{m}-\delta_{\mathrm{r}})$ and $\bm{a}_{\mathrm{r}}(\eta_{m}+\delta_{\mathrm{r}})$ are given as
\begin{eqnarray}
v_{n,m}^{\Delta}&=&\alpha\bm{a}^{*}_{\mathrm{r}}(\eta_{m}-\delta_{\mathrm{r}})\bm{a}_{\mathrm{r}}(\psi)\bm{a}^{*}_{\mathrm{t}}(\mu)\bm{a}_{\mathrm{t}}(\nu_{n}-\delta_{\mathrm{t}})x_{1}+\bm{a}^{*}_{\mathrm{r}}(\eta_{m}-\delta_{\mathrm{r}})\bm{n},\\
v_{n,m}^{\Sigma}&=&\alpha\bm{a}^{*}_{\mathrm{r}}(\eta_{m}+\delta_{\mathrm{r}})\bm{a}_{\mathrm{r}}(\psi)\bm{a}^{*}_{\mathrm{t}}(\mu)\bm{a}_{\mathrm{t}}(\nu_{n}+\delta_{\mathrm{t}})x_{1}+\bm{a}^{*}_{\mathrm{r}}(\eta_{m}+\delta_{\mathrm{r}})\bm{n}.
\end{eqnarray}
Assume $\bm{a}_{\mathrm{t}}(\nu_{n}+\delta_{\mathrm{t}})=\bm{a}_{\mathrm{t}}(\mu)$, the corresponding received signal strength of combining with $\bm{a}_{\mathrm{r}}(\eta_{m}-\delta_{\mathrm{r}})$ is calculated as
\begin{align}
\rho_{m}^{\Delta}=&\hspace{1mm}\left(v_{n,m}^{\Delta}\right)^{*}v_{n,m}^{\Delta}\\
=&\hspace{1mm}|\alpha|^{2}\bm{a}^{*}_{\mathrm{r}}(\eta_{m}-\delta_{\mathrm{r}})\bm{a}_{\mathrm{r}}(\psi)\bm{a}^{*}_{\mathrm{r}}(\psi)\bm{a}_{\mathrm{r}}(\eta_{m}-\delta_{\mathrm{r}})\nonumber\\
&+\bm{a}^{*}_{\mathrm{r}}(\eta_{m}-\delta_{\mathrm{r}})\bm{n}\bm{n}^{*}\bm{a}_{\mathrm{r}}(\eta_{m}-\delta_{\mathrm{r}})\nonumber\\
&+\alpha\bm{a}^{*}_{\mathrm{r}}(\eta_{m}-\delta_{\mathrm{r}})\bm{a}_{\mathrm{r}}(\psi)\bm{n}^{*}\bm{a}_{\mathrm{r}}(\eta_{m}-\delta_{\mathrm{r}})\nonumber\\
&+\alpha\bm{a}^{*}_{\mathrm{r}}(\eta_{m}-\delta_{\mathrm{r}})\bm{n}\bm{a}^{*}_{\mathrm{r}}(\psi)\bm{a}_{\mathrm{r}}(\eta_{m}-\delta_{\mathrm{r}}),
\end{align}
and the received signal strength of combining with $\bm{a}_{\mathrm{r}}(\eta_{m}+\delta_{\mathrm{r}})$ can be similarly obtained via $\rho_{m}^{\Sigma}=\left(v_{n,m}^{\Sigma}\right)^{*}v_{n,m}^{\Sigma}$. Regard $\rho_{m}^{\Delta}-\rho_{m}^{\Sigma}$ and $\rho_{m}^{\Delta}+\rho_{m}^{\Sigma}$ as difference and sum channel outputs. Denote by $S_{\Delta}$ the signal power of the difference channel output as
\begin{eqnarray}
S_{\Delta}=|\alpha|^{2}\left|\bm{a}^{*}_{\mathrm{r}}(\psi)\bm{\Lambda}_{\Delta}\bm{a}_{\mathrm{r}}(\psi)\right|.\label{diff}
\end{eqnarray}
Denote by $N_{\Sigma}$ the noise power of the sum channel output such that
\begin{eqnarray}
N_{\Sigma} = \sigma^2\left|\Upsilon_{\Sigma}\right|.\label{sum}
\end{eqnarray}
For unbiased estimator, i.e., $\mathbb{E}\left[\hat{\psi}_{m}\right]=0$ \cite{ap}, a reasonable approximation to the variance of angle estimation is expressed as \cite{handbook}
\begin{eqnarray}
\sigma_{\psi}^{2}=\mathbb{E}\left[\hat{\psi}_{m}^2\right]\approx\frac{1}{2k_{\mathrm{m}}^{2}S_{\Delta}/N_{\Sigma}}\left[1+\mathcal{M}^{2}(\psi)\right],\label{theoryvar}
\end{eqnarray}
where in this paper, $\mathcal{M}(\psi)=\zeta_{m}^{\mathrm{AoA}}$, and $k_{\mathrm{m}}=\mathcal{M}'(\eta_{m})$ is the slope of $\mathcal{M}(\cdot)$ at boresight of the corresponding auxiliary beam pair. According to (\ref{ori}),
\begin{equation}
k_{\mathrm{m}}=\mathcal{M}'(\eta_{m})=-\frac{\sin(\delta_{\mathrm{r}})}{1-\cos(\delta_{\mathrm{r}})}.\label{deriva}
\end{equation}
By plugging (\ref{diff}), (\ref{sum}) and (\ref{deriva}) into (\ref{theoryvar}), (\ref{varerror}) is obtained, which completes the proof.
\bibliographystyle{IEEEbib}
\bibliography{main_bib_WCL}
\end{document}